\documentclass[10pt,prb,twocolumn,notitlepage,showpacs,superscriptaddress]{revtex4-1}
\usepackage{amsthm,amsmath,amsfonts,amssymb,verbatim, color}
\usepackage{graphicx}
\usepackage{bm}
\usepackage{epsfig,slashed}
\usepackage[colorlinks=true,citecolor=blue,
linkcolor=blue,urlcolor=blue]{hyperref}
\begin{document}

\newcommand{\tr}{\mathop{\mathrm{Tr}}}
\newcommand{\bsigma}{\boldsymbol{\sigma}}
\newcommand{\re}{\mathop{\mathrm{Re}}}
\newcommand{\im}{\mathop{\mathrm{Im}}}
\renewcommand{\b}[1]{{\boldsymbol{#1}}}
\newcommand{\diag}{\mathrm{diag}}
\newcommand{\sign}{\mathrm{sign}}
\newcommand{\sgn}{\mathop{\mathrm{sgn}}}
\renewcommand{\c}[1]{\mathcal{#1}}

\newcommand{\mb}{\bm}
\newcommand{\ua}{\uparrow}
\newcommand{\da}{\downarrow}
\newcommand{\ra}{\rightarrow}
\newcommand{\la}{\leftarrow}
\newcommand{\mc}{\mathcal}
\newcommand{\bs}{\boldsymbol}
\newcommand{\lra}{\leftrightarrow}
\newcommand{\nn}{\nonumber}
\newcommand{\half}{{\textstyle{\frac{1}{2}}}}
\newcommand{\mf}{\mathfrak}
\newcommand{\MF}{\text{MF}}
\newcommand{\IR}{\text{IR}}
\newcommand{\UV}{\text{UV}}

\DeclareGraphicsExtensions{.png}

\title{Surface Majorana fermions and bulk collective modes in superfluid $^3$He-$B$}

\author{YeJe Park}
\email[electronic address: ]{yejepark@princeton.edu}
\affiliation{Department of Physics, Princeton University,
Princeton, New Jersey 08544, USA}

\author{Suk Bum Chung}
\affiliation{Center for Correlated Electron Systems, Institute for Basic Science (IBS), Seoul 151-747, Korea}
\affiliation{Department of Physics and Astronomy, Seoul National University, Seoul 151-747, Korea}
\affiliation{Department of Physics and Astronomy, University of California Los Angeles, Los Angeles, California 90095, USA}

\author{Joseph Maciejko}
\affiliation{Department of Physics, University of Alberta, Edmonton, Alberta T6G 2E1, Canada}
\affiliation{Princeton Center for Theoretical Science, Princeton University, Princeton, New Jersey 08544, USA}

\date\today

\begin{abstract}
The theoretical study of topological superfluids and superconductors has so far been carried out largely as a translation of the theory of noninteracting topological insulators into the superfluid language, whereby one replaces electrons by Bogoliubov quasiparticles and single-particle band Hamiltonians by Bogoliubov-de Gennes Hamiltonians. Band insulators and superfluids are, however, fundamentally different: while the former exist in the absence of inter-particle interactions, the latter are broken symmetry states that owe their very existence to such interactions. In particular, unlike the static energy gap of a band insulator, the gap in a superfluid is due to a dynamical order parameter that is subject to both thermal and quantum fluctuations. In this work, we explore the consequences of bulk quantum fluctuations of the order parameter in the $B$ phase of superfluid $^3$He on the topologically protected Majorana surface states. Neglecting the high-energy amplitude modes, we find that one of the three spin-orbit Goldstone modes in $^3$He-$B$ couples to the surface Majorana fermions. This coupling in turn induces an effective short-range two-body interaction between the Majorana fermions, with coupling constant inversely proportional to the strength of the nuclear dipole-dipole interaction in bulk $^3$He. A mean-field theory estimate of the value of this coupling suggests that the surface Majorana fermions in $^3$He-$B$ are in the vicinity of a quantum phase transition to a gapped time-reversal symmetry breaking phase.
\end{abstract}

\maketitle

\section{Introduction}

The prediction and discovery of time-reversal invariant topological band insulators---band insulators distinguished from their conventional counterparts by the existence of a bulk topological invariant and topologically protected edge or surface states, yet distinct from the time-reversal symmetry breaking quantum Hall insulator---is a major breakthrough in condensed matter physics.\cite{hasan2010,qi2011} The classification of such insulators requires only single-particle quantum mechanics, where interactions between electrons are ignored. Soon after the original predictions of the quantum spin Hall insulator and the three-dimensional (3D) topological insulator, it was realized that this topological band theory could be directly applied to the classification of Bogoliubov-de Gennes (BdG) Hamiltonians, which describe the spectrum of fermionic quasiparticles in paired superfluids and superconductors at the mean-field level. This led to the prediction of time-reversal invariant topological superfluids and superconductors.\cite{roy2006,roy2008,schnyder2008,qi2009,schnyder2009,kitaev2009} Translated in the superfluid or superconducting language, the surface states of topological band insulators become Majorana fermions---particles that are their own antiparticles, and which contain half the degrees of freedom of an ordinary complex fermion.\cite{wilczek2009} Under certain circumstances Majorana fermions possess non-Abelian statistics, which may lead to important applications in quantum information.\cite{nayak2008,alicea2012}

While the search for solid-state materials that exhibit topological superconductivity is still ongoing, a 3D topological superfluid has in principle already been found: the $B$ phase of superfluid $^3$He.\cite{VolovikBook} It was recognized early on by Salomaa and Volovik\cite{salomaa1988} that the spin-triplet $p$-wave order parameter in the Balian-Werthamer (BW) state\cite{balian1963,vdovin1963} that describes $^3$He-$B$ corresponds to a topologically nontrivial texture in momentum space, which in turn should give rise to protected fermionic zero modes at the boundary of the sample.\cite{volovik2009}

Although the translation of topological band theory into the superfluid/superconducting context has led to remarkable predictions and insights, superfluids and superconductors remain fundamentally distinct from band insulators. While in the latter inter-particle interactions can be treated as a perturbation on top of the noninteracting band structure, the former are broken symmetry states that owe their very existence to such interactions. Unlike the frozen energy gap of a band insulator, the gap in a superfluid or superconductor originates from a dynamical order parameter that fluctuates even at zero temperature.

In this work, we go beyond the pure BdG description of topological superfluids that has been the focus of much work in this field to date, and explore the consequences of bulk order parameter fluctuations in the only known 3D topological superfluid, $^3$He-$B$. In particular, we are interested in the question of how the properties of the Majorana surface states in $^3$He-$B$ are affected by such fluctuations. The fluctuations that are likely to have the most impact are the gapless Goldstone modes of $^3$He-$B$, while fluctuations in the amplitude of the order parameter have a gap on the order of the bulk energy gap and can be neglected at the energy scale of the surface states. Related work by Grover and Vishwanath\cite{grover2012} studied the coupling between Majorana surface states and bulk fluctuations in topological superconductors. However, the bulk fluctuations they consider are fluctuations of a magnetic order parameter that is assumed to exist in addition to the physics of superfluidity, while we are considering fluctuations of the superfluid order parameter itself (which gives rise to Majorana surface states in the first place). In other words, the physics we focus on is intrinsic to superfluidity in $^3$He-$B$ and does not require the proximity to a novel quantum critical point. Other conceptually related work includes the study of thermal fluctuations of the order parameter in 3D $p$-wave superconductors\cite{li2009} and 2D chiral topological superconductors,\cite{bauer2013} as well as the study of proximity-induced topological superconductivity by a 1D superconductor where quantum fluctuations imply algebraically decaying superconducting correlations but no true long-range order.\cite{fidkowski2011}

\begin{figure}
\includegraphics[width=1\linewidth]{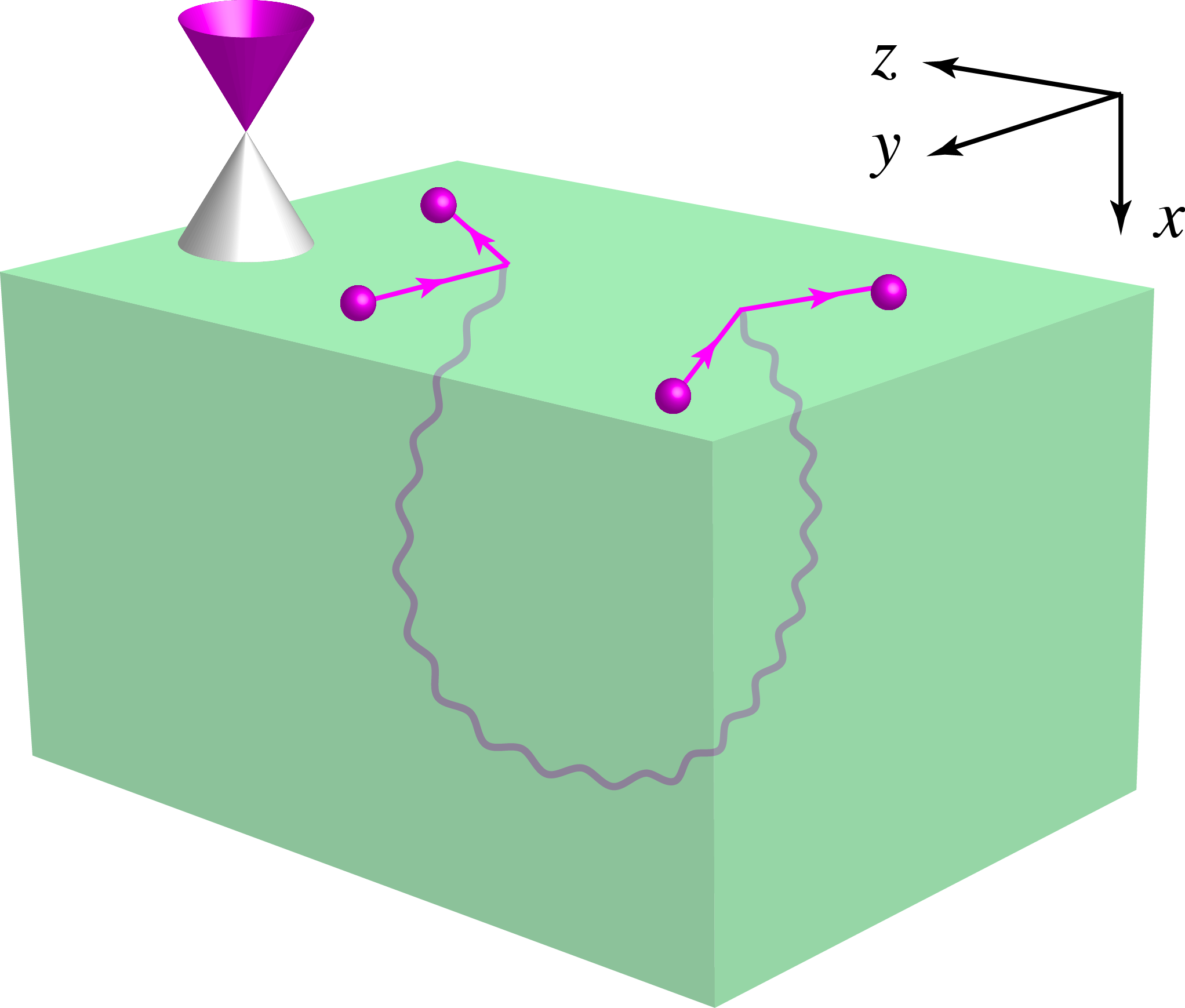}
\caption{Majorana fermions (magenta spheres) on the surface of $^3$He-$B$ with the energy-momentum dispersion of a cone (magenta cone; the negative-energy part of the spectrum illustrated in white is redundant) can effectively interact by exchanging quanta of the bulk collective modes (wiggly line).}
\label{fig:surfbulk}
\end{figure}

Our main results may be summarized as follows. Out of the four Goldstone modes in $^3$He-$B$---the phase mode and the three spin-orbit modes---we find that only one spin-orbit mode couples to the surface Majorana fermions. Unlike the phase mode, the spin-orbit modes are in fact not truly gapless: they acquire a small gap due to the dipole-dipole interaction between nuclei of the $^3$He atoms. Nevertheless, the surface Majorana fermions can exchange quanta of this bulk mode, leading to an effective short-range four-fermion interaction between them (Fig.~\ref{fig:surfbulk}) with a coupling constant that is inversely proportional to the strength of the dipole-dipole interaction. This interaction is perturbatively irrelevant in the renormalization group sense, but can lead to a quantum phase transition towards a gapped surface phase with spontaneously broken time-reversal symmetry if the coupling constant exceeds a certain critical value. We estimate this critical value in mean-field theory, and find that the effective surface coupling constant in $^3$He-$B$ is of the same order of magnitude as the critical one. This suggests that the Majorana surface states in $^3$He-$B$ are in the vicinity of a quantum phase transition to a time-reversal symmetry breaking phase, if not already in that phase. This latter possibility is not necessarily excluded by recent experimental work consistent with the presence of gapless surface states in $^3$He-$B$, because a small but nonzero gap would be hard to detect. Although our mean-field theory predicts that the transition is first order, it could ultimately become continuous if fluctuation effects are considered, in which case the corresponding quantum critical point should exhibit an emergent $\c{N}=1$ supersymmetry.\cite{ScottThomas,sonoda2011,grover2012,grover2014}

The strategy we adopt in this paper is as follows. We begin by reviewing how to solve for the Majorana fermion surface modes in a static order parameter background (Sec.~\ref{sec:MajSS}). We then allow for small fluctuations of the order parameter, and determine how these couple to the fermionic surface modes (Sec.~\ref{sec:SBcoupling}). Integrating out the bulk order parameter fluctuations, we derive an effective surface interaction between the Majorana fermions (Sec.~\ref{sec:EffSurfInt}). Finally, we use mean-field theory to determine possible broken symmetry states induced by this interaction (Sec.~\ref{sec:BrokenSymmetry}).

\section{Majorana surface states of $\,^3\text{He}$-$B$}
\label{sec:MajSS}

We begin by reviewing the derivation of the Majorana fermion surface states from the BdG mean-field description of the $^3$He-$B$ superfluid (see, e.g., Ref.~\onlinecite{okuda2012} and references therein). We denote the annihilation (creation) operator for a neutral $\,^3$He fermionic quasiparticle by $c_{\b{k}\sigma}$ ($c_{\b{k}\sigma}^\dag$) where $\sigma = \ua, \, \da$ is the spin quantum number and $\b{k}$ is the 3D spatial momentum quantum number, and use units such that $\hbar=1$. When the neutral fermions are in the $\,^3$He-$B$ superfluid phase, the system is described by the time-reversal invariant Hamiltonian,
\begin{equation}
H_B = \sum_{\b{k}} \Psi^{\dagger}_{\b{k}} \mc H_{\text{BdG}}(\b{k}) \Psi_{\b{k}}, \label{bulk}
\end{equation}
where the Nambu spinor $\Psi_{\b{k}}$ is defined as
\begin{align}
\Psi_{\b{k}} &=
\begin{pmatrix}
c_{\b{k} \ua} & c_{\b{k}\da} & c^{\dagger}_{-\b{k}\da} & -c_{-\b{k} \ua}^{\dagger}
\end{pmatrix}^T
=
\begin{pmatrix}
c_{\b{k}\sigma} \\ i\sigma^y_{\sigma\sigma'} c^{\dagger}_{-\b{k}\sigma'}
\end{pmatrix}, \label{Nambu}
\end{align}
and the spin-triplet $p$-wave pairing BdG Hamiltonian $\mc H_{\text{BdG}}(\b{k})$ is defined as
\begin{equation}
\mc H_{\text{BdG}}(\b{k}) =
\begin{pmatrix}
\epsilon_{\b{k}}  & (\Delta_0/k_F)   \sigma^{\mu} R_{\mu j}  k_j\\
(\Delta_0/k_F)   \sigma^{\mu}  R_{\mu j} k_j & -\epsilon_{\b{k}}
\end{pmatrix},\label{BdG}
\end{equation}
corresponding to the BW state.\cite{balian1963,vdovin1963}
Here, $\epsilon_{\b{k}} = \b{k}^2/2m - E_F$ where $E_F=k_F^2/2m$ is the Fermi energy in the normal state of $\,^3$He, $m$ is the effective mass of the fermionic quasiparticles, $k_F$ is the Fermi momentum, and $\Delta_0$ is the energy gap (that can be made real by a uniform gauge transformation). $R_{\mu j}$ is a constant $SO(3)$ relative rotation matrix\cite{VollhardtWolfle}  that relates the spin coordinate system indexed by $\mu = x,y,z$ and the spatial coordinate system indexed by $j = x,y,z$. The corresponding relative $SO(3)$ rotation group is conventionally denoted by $SO(3)_{L-S}$. We denote the usual Pauli matrices by $\sigma^{\mu}=(\sigma^x,\sigma^y,\sigma^z)$. The single-particle excitations in the bulk are the gapped Bogoliubov quasiparticles with isotropic energy dispersion $E(\b{k}) = \sqrt{\epsilon_{\b{k}}^2+ \Delta_0^2}$.

A generic relative rotation matrix $R_{\mu j}$ may be parameterized  by the rotation axis $\hat{\b n}$ and angle of rotation $\theta$,
\begin{equation}\label{RotMatrix}
  R_{\mu j}( \hat{\b n},\theta) = (1-\cos \theta)\hat n_{\mu}\hat n_j +\delta_{\mu j}\cos \theta - \epsilon_{\mu j k} \hat n_k \sin \theta,
\end{equation}
where $\hat{\b{n}}^2=1$. Each relative rotation matrix $R_{\mu j }( \hat{\b n},\theta)$ represents one member of a family of possible BW states. If the nuclear spin of the $^3$He atoms is neglected, these states are all degenerate in energy. In reality, the dipole-dipole interaction between the nuclear spins of the $\,^3$He atoms leads to a specific value of $\theta$ being energetically favored, the so-called Leggett angle $\theta_L = \cos^{-1}\left(-\frac{1}{4}\right)$.\cite{leggett1973,leggett1974,brinkman1974b} The remaining parameter $\hat{\b n}$ remains free in the bulk, but the dipole-dipole interaction in the presence of a surface with normal $\hat {\b s}$ tends to align $\hat{\b n}$ along $\hat{\b s}$ in the vicinity of the surface within the coherence length $\xi_0\sim v_F/\Delta_0$\cite{smith1977} (which characterizes the extension of a Cooper pair) with $v_F=k_F/m$ the Fermi velocity in the normal state of $^3$He. For our purposes, the effect of the surface on $\hat {\b n}$ may be treated as a boundary condition on $\hat{\b n}$. We consider a volume of $\,^3$He-$B$ superfluid occupying a semi-infinite 3D region $x >0$ with a 2D flat surface corresponding to the $yz$ plane, and the normal is $\hat {\b s} =- \hat {\b x}$ (Fig.~\ref{fig:surfbulk}). Given that $\hat{\b{n}}$ is free in the bulk, without loss of generality we may choose $\hat{\b{n}}=-\hat{\b{x}}$ as our reference equilibrium state in the bulk. The corresponding relative rotation matrix $R^{(0)}_{\mu j}$ is then
\begin{equation}
  R_{\mu j}^{(0)} =
  \begin{pmatrix}
    1 & 0 & 0\\
    0  &\cos\theta_L & \sin\theta_L\\
    0 &- \sin\theta_L & \cos \theta_L
  \end{pmatrix}. \label{R_0}
\end{equation}

\subsection{Majorana surface states}

In the presence of a surface, there exist fermionic modes (Andreev bound states) localized at this surface with energies within the bulk gap. As explained in the introduction, we will first solve for the wave function and spectrum of these modes in the static order parameter background Eq.~(\ref{R_0}), and then allow for small order parameter fluctuations above the background. In first quantization, the BdG Hamiltonian (\ref{BdG}) becomes
\begin{equation}
 \hat H =
\begin{pmatrix}
\hat p^2/2m- E_F  & (\Delta_0/k_F)  \bsigma \cdot \hat{\mb p} \\
(\Delta_0/k_F ) \bsigma \cdot  \hat{\mb p}& - \hat p^2/2m+ E_F
\end{pmatrix}\label{first_quant},
\end{equation}
where we use the caret ($\hat\,$) to denote first-quantized operators ($\hat{\b{p}}=-i\nabla$). The dependence of the Hamiltonian on the Leggett angle $\theta_L$ via Eq.~(\ref{R_0}) has been eliminated by a rotation of the spatial coordinates relative to the spin coordinates in the $yz$ plane by the angle $\theta_L$, so that $k_y\cos\theta_L + k_z\sin\theta_L \ra k_y$ and $- k_y\sin\theta_L+k_z\cos\theta_L \ra k_z$. The surface states are the solutions of the time-independent Schr\"odinger equation for this Hamiltonian,
\begin{equation}
E \phi (\mb r)=  \hat H \phi(\mb r), \label{Schrodinger_eq}
\end{equation}
where $\b{r}=(x,y,z)$, and we assume the Dirichlet boundary conditions $\phi(0,y,z)=0$ and $\phi(\infty,y,z) = 0$. Although the details of the wave function of the surface states will depend on the type of boundary conditions, the existence of the surface states will not, because of their topological character.\cite{schnyder2008} We consider an ansatz of the form
\begin{subequations}
\begin{align}
\phi(\mb  r)& = \psi_{\mb k_{\parallel},\pm}(\mb r)  \phi_0, \\
\psi_{\mb k_{\parallel},\pm}(\mb r) & = \mc N e^{i \mb k_{\parallel}\cdot \mb r_\parallel} e^{\pm i k_{\perp} x}\, \chi(x),\label{PsiAnsatz}
\end{align}
\end{subequations}
where $\mc N$ is a normalization constant, $\mb k_{\parallel} = (k_y,k_z) = (k_1,k_2)$ and $\b{r}_\parallel=(y,z)$, $k_{\perp} = \sqrt{k_F^2 - |\mb k_{\parallel}|^2}$,  $\chi(x)$ is a scalar function of $x$, and $\phi_0$ is a 4D spinor. In the weak-pairing limit,\cite{nagato2009,chung2009}
\begin{equation}
 k_{\perp}\gg  \kappa\equiv k_F \frac{\Delta_0}{E_F} , \label{weak-pair}
\end{equation}
the substitution of the ansatz into Eq.~(\ref{Schrodinger_eq}) gives
\begin{equation}\label{SchrodEq8}
E \chi(x) \phi_0 =\left( H_0(\mb k_{\parallel}) \pm \hat H_{\perp}\right) \chi(x) \phi_0 ,
\end{equation}
where
\begin{subequations}
\begin{align}
H_0(\mb k_{\parallel})&
 = \begin{pmatrix}
0 & (\Delta_0/k_F) \mb k_{||}\cdot\bs\sigma \\
(\Delta_0/k_F) \mb k_{||}\cdot\bs\sigma & 0
 \end{pmatrix},\label{Eq9a} \\
\hat H_{\perp}&
= k_{\perp}
\begin{pmatrix}
(1/m)(-i \partial_x) & (\Delta_0/k_F)  \sigma^x \\
(\Delta_0/k_F)\sigma^x &-(1/m)(-i \partial_x)
 \end{pmatrix}.\label{Eq9b}
\end{align}
\end{subequations}
The gapless surface states are eigenstates of the operator $\hat H_{\perp}$ with eigenvalue zero, since then Eq.~(\ref{SchrodEq8}), (\ref{Eq9a}) and (\ref{Eq9b}) imply that $E=0$ at $\b{k}_\parallel=0$. This condition is satisfied by choosing two independent spinors $\phi^{\sigma}_0$,
\begin{subequations}
\begin{align}
\phi_0^{\ua}&  = e^{-i\pi/4}
\begin{pmatrix}
1 & 0 &0 & -i
\end{pmatrix}^T,\label{phi0up}\\
\phi_0^{\da}&  = e^{i\pi/4}
\begin{pmatrix}
0 & 1 &-i & 0
\end{pmatrix}^T,\label{phi0down}
\end{align}
\end{subequations}
as well as
\begin{equation}\label{chiofx}
\chi(x) = e^{-\kappa  x/2},
\end{equation}
which manifestly satisfies the Dirichlet boundary condition at $x=\infty$. The surface states are labeled by the surface momentum $\mb k_{\parallel}$ and the spin index $\sigma$. Considering Eqs.~(\ref{PsiAnsatz}), (\ref{weak-pair}), and (\ref{chiofx}), we see that the weak-pairing limit corresponds to BdG wave functions $\phi(\b{r})$ that only involve momenta near the Fermi surface.

The solution that satisfies the Dirichlet boundary condition at $x = 0$ is given by a linear superposition of $\psi_{\mb k_{\parallel} ,+}(\mb r)$ and $\psi_{\mb k_{\parallel} ,-}(\mb r)$,
\begin{subequations}
\begin{align}
\phi^{\sigma}(\mb r) & = \psi_{\mb k_{\parallel}}(\mb r) \phi_0^{\sigma},\\
\psi_{\mb k_{\parallel}}(\mb r) & = \mc N e^{i \mb k_{\parallel}\cdot \mb r_\parallel} \sin(k_{\perp}x)\theta(x)\, \chi(x),
\end{align}
\end{subequations}
where we explicitly include the Heaviside step function $\theta(x)$ to signify that the superfluid occupies the $x>0$ half-space.

The normalization constant $\mathcal{N}$ remains to be determined. This is most easily done by considering a finite volume $V=L_{\parallel}^2 L_{\perp}$ of superfluid of length $L_\parallel$ in the $y$ and $z$ directions and $L_\perp$ in the $x$ direction. In general, $\mathcal{N}$ depends on the magnitude of $\mb k_{\parallel}$, but in the weak-pairing limit (\ref{weak-pair}) and in the limit of large system size $L_{\perp} \gg \kappa^{-1}$ (such that it is meaningful to have $\phi$ vanish at $x=\infty$ even though the system has a finite extent in the $x$ direction), we find $\mc N =  L_{\parallel}^{-1} \sqrt{2 \kappa}$, and
\begin{equation}
\int_{x>0} d^3 \mb r \, |\psi_{\mb k_{\parallel}}(\mb r)|^2 = 1.
\end{equation}

As we will be considering interaction effects among the surface states, it is convenient to describe them in second quantization. The fermionic field operator $\hat \psi_{\sigma}(\mb r)$ can be expanded as
\begin{equation}
\hat \psi_{\sigma}(\mb r) = \sum_{\mb k_{\parallel}} \psi_{\mb k_{\parallel}}(\mb r) c_{\mb k_{\parallel}\sigma} + \dots, \label{field}
\end{equation}
where $c_{\mb k_{\parallel}\sigma}$ annihilates a fermion with spatial wave function $\psi_{\mb k_{\parallel}}(\mb r)$ and spin $\sigma = \ua,\da$. The extra terms ($\ldots$) are associated with gapped bulk modes. The field operator satisfies the usual anticommutation relations,
\begin{equation}\label{anticommutation}
\{\hat \psi_{\sigma}(\mb r),\hat \psi_{\sigma'}^\dag(\mb r')\}
=\delta_{\sigma\sigma'} \delta^{(3)}(\mb r - \mb r').
\end{equation}
The two orthogonal spinors $\phi^{\ua}_0,\phi^{\da}_0$ with eigenvalue zero in Eq.~(\ref{phi0up})-(\ref{phi0down}) are associated with two gapless fermionic modes $\gamma_{\b{k}_\parallel\ua},\gamma_{\b{k}_\parallel\da}$. Given that the spinor part of the Hilbert space on which $\hat{H}_\perp$ in Eq.~(\ref{Eq9b}) acts is four-dimensional, there are two other orthogonal spinors with nonzero eigenvalue that correspond to gapped modes $\bar \gamma_{\mb k_{\parallel}\ua},\bar \gamma_{\mb k_{\parallel}\da}$. The microscopic $^3$He quasiparticle operators $c_{\mb k_{\parallel}\ua},c_{\mb k_{\parallel}\da}$ are linear combinations of both gapless and gapped modes,
\begin{subequations}
\begin{align}
&c_{\mb k_{||}\ua} = \frac{1}{\sqrt 2}\left(e^{i\pi/4} \gamma_{\mb k_{||}\ua}+e^{-i\pi/4}\bar \gamma_{\mb k_{||} \ua}\right),\\
&c_{\mb k_{||}\da} = \frac{1}{\sqrt 2} \left(e^{-i\pi/4}\gamma_{\mb k_{||}\da}+e^{i\pi/4}\bar \gamma_{\mb k_{||}\da} \right),
\end{align}
\end{subequations}
while $\gamma_{\mb k_{\parallel}\sigma}$ itself is a linear combination of $c_{\mb k_{\parallel}\sigma}$ and $c_{\mb k_{\parallel}\sigma}^{\dagger}$, \begin{subequations}
\begin{align}
\gamma_{\mb k_{\parallel} \ua} & = \frac{e^{-i\pi/4}}{\sqrt 2} \left(c_{\mb k_{\parallel} \ua} + i c_{-\mb k_{\parallel} \ua}^{\dagger}\right),\\
\gamma_{\mb k_{\parallel} \da} & = \frac{e^{i\pi/4}}{\sqrt 2} \left(c_{\mb k_{\parallel} \da} -i c_{-\mb k_{\parallel} \da}^{\dagger}\right).
\end{align}
\end{subequations}
The gapless modes $\gamma_{\mb k_{\parallel} \sigma}$ are known as Majorana fermion operators because they satisfy the reality condition
\begin{align}\label{reality}
\gamma_{\b{k}_\parallel\sigma}^\dag=\gamma_{-\b{k}_\parallel\sigma},
\end{align}
or, equivalently, the Clifford algebra
\begin{equation}
\{\gamma_{\mb k_{\parallel} \sigma},\gamma_{-\mb k'_{\parallel} \sigma'}\}  = \delta_{\mb k_{\parallel},\mb k'_{\parallel}}^{(2)} \delta_{\sigma,\sigma'}.
\end{equation}
In the low-energy limit, we can neglect the gapped modes $\bar{\gamma}_{\b{k}_\parallel\sigma}$ and approximate the full field operator by
\begin{equation}
\hat \psi_{\sigma}(\mb r) \approx\frac{1}{\sqrt 2}\sum_{\mb k_{\parallel}\sigma'} e^{i \pi\sigma^z_{\sigma\sigma'}/4}\psi_{\mb k_{\parallel}}(\mb r)  \gamma_{\mb k_{\parallel}\sigma'}. \label{field_op_exp}
\end{equation}
We can now write down a second-quantized Hamiltonian for the noninteracting Majorana surface states. Given that $H_0(\mb k_{\parallel})$ in Eq.~(\ref{Eq9a}) is effectively a Hamiltonian matrix for the surface states, we have
\begin{equation}\label{H0Majorana}
H_0 = \frac{\Delta_0}{2k_F} \sum_{\mb k_{\parallel} }\gamma^{T}_{-\mb k_{\parallel} } (\mb k_{\parallel}\cdot \tilde{\bs \sigma}) \gamma_{\mb k_{\parallel} },
\end{equation}
where it is convenient to define rotated Pauli matrices $\tilde \sigma^{\mu}$ due to the phase factors in Eq.~(\ref{field_op_exp}),
\begin{subequations}
\begin{align}
&\tilde \sigma^y = \tilde \sigma^1 = \sigma^z, \\
&\tilde \sigma^z = \tilde \sigma^2 = \sigma^x.
\end{align}
\end{subequations}
The Hamiltonian (\ref{H0Majorana}) has a cone-like linear dispersion
\begin{equation}
E(\mb k_{\parallel}) = \Delta_0\frac{|\mb k_{\parallel}|}{k_F}.
\end{equation}
We ignore negative eigenenergies that do not correspond to physical states but simply arise from the particle-hole redundancy of the BdG description.

\section{Surface interactions mediated by bulk Goldstone modes}

The derivation of the surface states in the previous section assumed a static bulk order parameter with constant and uniform pairing amplitude $\Delta_0$ and relative rotation matrix $R^{(0)}_{\mu j}$ [Eq.~(\ref{R_0})]. In a real helium sample however, the order parameter is a dynamical field that fluctuates even at zero temperature due to quantum zero-point motion. The quanta of this dynamical field can be absorbed and emitted by the surface Majorana fermions, and can thus mediate interactions between the Majorana fermions. The purpose of this section is to derive the form of these interactions. In a first stage, we determine the form of the coupling between the Majorana surface states and the fluctuations of the bulk order parameter, i.e., the bulk collective modes. In a second stage, we integrate out these bulk collective modes to derive the form of the intra-surface interactions. Although we will focus on a semi-infinite geometry with a single surface that is a good approximation for a thick helium sample, a similar calculation could be performed in a slab geometry that would describe helium thin films---although the film should not be so thin that the $A$ phase is favored over the $B$ phase.\cite{vorontsov2003} In this case there would also be inter-surface interactions where a bulk order parameter fluctuation is emitted by a Majorana fermion on the (say) top surface, propagates through the bulk to the bottom surface, and is absorbed by a Majorana fermion on that surface.

There are numerous collective modes in the $B$ phase of $^3$He. This phase spontaneously breaks the $SO(3)_L\times SO(3)_S\times U(1)_N$ symmetry of the parent Fermi liquid state to $SO(3)_{L+S}$, where $SO(3)_L$ and $SO(3)_S$ correspond to spatial and spin rotations, respectively, $U(1)_N$ describes particle number conservation, and $SO(3)_{L+S}$ describes simultaneous rotations in real space and spin space.\cite{VollhardtWolfle} If we ignore the dipole-dipole interaction, the associated Goldstone manifold is $SO(3)_{L-S}\times U(1)_N$, corresponding to relative rotations in real space and spin space as well as phase rotations. As a result there are four gapless Goldstone modes in $^3$He-$B$: three spin-orbit modes\cite{brinkman1974b} and one phase mode. As we will see however, the dipole-dipole interaction generates a small gap for some of the gapless spin-orbit modes. Furthermore, there are also gapped amplitude modes,\cite{wolfle1977} but these have energies of the order of the bulk gap and can be ignored in a first approximation.

\subsection{Surface-bulk coupling}
\label{sec:SBcoupling}

In Sec.~\ref{sec:MajSS}, the equilibrium value of the bulk $p$-wave pairing order parameter was chosen to be
\begin{equation}
\Delta(\b{k}) =  \frac{\Delta_0}{k_F} \sigma^{\mu} i\sigma^y R^{(0)}_{\mu j }k_j.
\end{equation}
As done in our derivation of the Majorana surface states, we can rotate the spatial coordinates on the surface plane so that the Leggett angle is eliminated, and the order parameter becomes
\begin{equation}
\Delta(\b{k}) =  \frac{\Delta_0}{k_F} \sigma^{\mu}i \sigma^y \delta_{\mu j }k_j.
\end{equation}
We now include the effect of the gapless fluctuations of the order parameter, i.e., the bulk Goldstone modes. These correspond to small variations of the relative rotation matrix $R_{\mu j}(\mb R)$ and the real phase $\varphi(\mb R)$,
\begin{equation}
\Delta(\b{k} ; \mb R) \simeq \frac{\Delta_0}{k_F} \left(1 + i\varphi(\mb R)\right) \sigma^{\mu} i\sigma^y R_{\mu j}(\mb R) k_j, \label{OP}
\end{equation}
where $\b{k}$ is the relative momentum of the fermion pair, the position vector $\mb R$ is the center of mass (CM) of the pair, and we consider small fluctuations $\varphi(\b{R})\ll 2\pi$ (i.e., we only consider vortex-free field configurations). The fluctuations occur on a length scale much larger than $k_F^{-1}$, while the magnitude of the relative momentum $\b{k}$ of the pair is of order $k_F$. The relative rotation matrix can be expanded in terms of the three independent generators of $SO(3)_{L-S}$,
\begin{equation}
S^{(\alpha)}_{\mu \nu} = -i \epsilon_{\alpha\mu \nu},
\end{equation}
where $\epsilon_{\alpha\mu \nu}$ is the Levi-Civita antisymmetric tensor and $\alpha, \mu ,\nu = x,y,z$. The spin-orbit fluctuations are parameterized by three real bosonic fields $\theta_{\alpha}(\b{R})$, hence we have
\begin{equation}
 R_{\mu j} (\mb R) \simeq \left(\delta_{\mu\nu}+ i\theta_{\alpha}(\mb R) S^{(\alpha)}_{\mu\nu}\right)\delta_{\nu j}, \label{small_order}
\end{equation}
where $\theta_\alpha(\b{R})\ll 2\pi$ here also.

The coupling of the order parameter with the Bogoliubov quasiparticles can be obtained from the bulk BdG Hamiltonian (\ref{bulk}), generalized to include CM degrees of freedom,
\begin{equation}
H_{\text{coupling}} = \frac{1}{2V}\sum_{\b{k},\b{Q}} c_{\b{k}+\b{Q}/2,\sigma}^{\dagger} c^{\dagger}_{-\b{k}+\b{Q}/2, \sigma'}\Delta_{\sigma\sigma'}(\b{k};\b{Q})
+\mathrm{H.c.}, \label{coupling0}
\end{equation}
where $\b{Q}$ is the CM momentum obtained by Fourier transforming with respect to  $\mb R$, and the fermion operators $c_{\mb k\sigma}$ are the Fourier transforms of the field operators $\hat \psi_{\sigma}(\mb r)$ in Eq.~(\ref{field_op_exp}),
\begin{align}
c_{\mb k\sigma}& = \frac{1}{V^{1/2}} \int d^3\mb r \,e^{-i\mb k\cdot \mb r} \hat \psi_{\sigma}(\mb r) \nonumber \\
& = \psi(k_x) c_{\mb k_{\parallel}\sigma} + \dots,
\end{align}
where
\begin{align}
\psi(k_x) & = \sqrt{\frac{2\kappa}{L_{\perp} }} \int_0^\infty dx \, e^{-i k_x x} \sin(k_{\perp} x) \chi(x)  \nonumber \\
& = \sqrt{\frac{2\kappa}{ L_{\perp}}}\frac{k_{\perp}}{k_{\perp}^2-(k_x -i\kappa/2 )^2}
\end{align}
is an envelope function that describes the finite penetration depth $\propto\kappa^{-1}\sim\xi_0$ of the Majorana surface states into the bulk.

Inserting into Eq.~(\ref{coupling0}) the order parameter given in Eq.~(\ref{OP}) and Eq.~(\ref{small_order}), and discarding the gapped modes $\bar \gamma_{\mb k_{\parallel}\sigma}$, we obtain
\begin{align}
H_{\text{coupling}} = \frac{\Delta_0}{2V} &\sum_{\b{Q}}\Bigl[
 -i \varphi(-\b{Q})\delta_{\mu j} \left(J_j^{\mu}(\b{Q}) - J_j^{\mu}(-\b{Q})^{\dagger}\right) \nonumber \\
 +&i\theta_{\alpha}(-\b{Q}) S^{(\alpha)}_{\mu j} \left(J_j^{\mu}(\b{Q}) + J_j^{\mu}(-\b{Q})^{\dagger}\right)\Bigr], \label{coupling}
\end{align}
where the quantities $J_j^{\mu}(\b{Q})$ are defined as
\begin{align}
J_j^{\mu}(\b{Q})
= \frac{1}{2k_F} &\sum_{\b{k}} k_j \psi(-k_x + Q_x/2)\psi(k_x+Q_x/2) \nonumber\\
 \times& \gamma_{-\mb k_{\parallel} + \mb Q_{\parallel}}^T(-i \sigma^y e^{i\pi \sigma_z/4} \sigma^{\mu}e^{i\pi\sigma_z/4} )\gamma_{\mb k_{\parallel} + \mb Q_{\parallel}}. \label{J}
\end{align}
Taking the Hermitian conjugate of $J_j^{\mu}(\b{Q})$, we find that  $J_j^x(\mb R)$ is anti-Hermitian while $J_j^y(\mb R)$ and $J_j^z(\mb R)$ are Hermitian,
\begin{subequations}
\begin{align}
 J_j^{x}(\b{Q})^{\dagger} &= -J_j^x(-\b{Q}), \\
 J_j^y(\b{Q})^{\dagger} &= J_j^y(-\b{Q}),\label{Jyj} \\
 J_j^z(\b{Q})^{\dagger}& = J_j^z(-\b{Q}).\label{Jzj}
 \end{align}
\end{subequations}
The summand in Eq.~(\ref{J}) for $j = x$ is odd under $k_x \rightarrow -k_x$, thus in fact $J_x^{\mu}(\b{Q})$ vanishes identically for all $\mu$.

From Eq.~(\ref{coupling}), we see that the phase fluctuation $\varphi(\b{R})$ couples linearly to a Hermitian operator $\mathcal{O}_\varphi(\b{R})$ with Fourier transform
\begin{align}
\mathcal{O}_\varphi(\b{Q})=-i\left(J^\mu_\mu(\b{Q})-J^\mu_\mu(-\b{Q})^\dag\right),
\end{align}
which vanishes identically because $J^x_x(\b{Q})=0$ and because of Eq.~(\ref{Jyj})-(\ref{Jzj}). Therefore there is no coupling between surface Majorana fermions and phase fluctuations. Likewise, the spin-orbit fluctuations $\theta_\alpha(\b{R})$ couple linearly to Hermitian operators $\mathcal{O}_{\theta_\alpha}(\b{R})$ with Fourier transform
\begin{align}
\mathcal{O}_{\theta_\alpha}(\b{Q})=\epsilon_{\alpha\mu j}\left(J_j^{\mu}(\b{Q}) + J_j^{\mu}(-\b{Q})^{\dagger}\right).
\end{align}
Since $\epsilon_{\alpha\mu j}$ is antisymmetric under $j\leftrightarrow \mu$ and $J_x^{\mu}(\b{Q}) = 0$ for all $\mu$, the only possibility is that $\theta_{x}(-\mb Q)$ couples to $J^y_z(\mb Q)$ and $J_y^z(\mb Q)$.

We therefore obtain the coupling between bulk Goldstone modes and surface Majorana fermions $\gamma_{\mb k_{\parallel}\sigma}$ as
\begin{equation}
H_{\text{coupling}} = \frac{\Delta_0}{V} \sum_{\b{Q}} \theta_x(-\b{Q}) \rho (\b{Q}),\label{HcouplingThetax}
\end{equation}
where we define the Majorana bilinear
\begin{align}
\rho(\b{Q})& = J_z^y(\b{Q}) - J_y^z(\b{Q})\nonumber\\
& = \sum_{k_x}\psi(-k_x+Q_x/2)\psi(k_x+ Q_x/2)  \rho(\mb Q_{\parallel}), \label{rho}
\end{align}
where
\begin{align}
\rho(\mb Q_{\parallel})=\frac{1}{2k_F}\sum_{\mb k_{\parallel}}\gamma_{-\mb k_{\parallel} + \mb Q_{\parallel}/2}^T[\hat{\mb x}\cdot(\mb k_{\parallel}\times \tilde{\bs \sigma})]\gamma_{\mb k_{\parallel}+ \mb Q_{\parallel}/2}.
\end{align}
Performing the summation over $k_x$ in Eq.~(\ref{rho}), we obtain
\begin{equation}
\sum_{k_x} \psi(-k_x + Q_x/2) \psi(k_x + Q_x/2)
=\frac{1}{ 1+i( Q_x/\kappa)},
\end{equation}
in the weak-pairing limit (\ref{weak-pair}) and assuming that the CM momentum $Q_x$ is small compared to $k_\perp$.

We note that the coupling (\ref{HcouplingThetax}) between the Goldstone mode $\theta_x$ and the Majorana bilinear $\rho$ does not vanish at $\b{Q}=0$. In the bulk of a superfluid, or any ordered state with a spontaneously broken continuous global symmetry, the coupling of a Goldstone mode with other degrees of freedom such as fermionic quasiparticles typically vanishes at the ordering wave vector (here $\b{Q}=0$), a general result first obtained by Adler.\cite{adler1965a,*adler1965b} Interactions between Goldstone modes and other degrees of freedom can only occur through derivative couplings, to preserve the invariance of the low-energy effective action under uniform rotations within the Goldstone manifold. Here the coupling (\ref{HcouplingThetax}) does not vanish at $\b{Q}=0$ because the Majorana fermions, being localized in real space at the sample surface, are a linear superposition of all bulk momentum eigenstates. Scattering of a surface Majorana fermion by a bulk Goldstone boson generally involves large bulk momentum transfers, a consequence of the explicit breaking of translation symmetry by the sample surface, and Adler's principle does not apply. A more straightforward way to see why the coupling between $\theta_x(\bm Q=\bm 0)$ and the Majorana fermions does not vanish is to note that a coupling of this type can be generated by a uniform global rotation in spin space around the $x$ axis (surface normal) by an infinitesimal angle $\theta_x$, i.e., $\tilde{\sigma}_i\rightarrow\tilde{\sigma}_i+\theta_x\epsilon_{ij}\tilde{\sigma}_j$.

In summary, the only fluctuation of the bulk order parameter that couples to the surface Majorana fermions is the spin-orbit mode $\theta_x$. That $\theta_y$ and $\theta_z$ do not couple at all reflects the anisotropy of the spin susceptibility characteristic of the surface Majorana fermions.\cite{chung2009,nagato2009} The absence of coupling to the phase fluctuation $\varphi$ can be understood from the charge neutrality of Majorana fermions.

\subsection{Effective surface interactions}
\label{sec:EffSurfInt}

Effective interactions between the surface Majorana fermions can be derived by integrating out the bulk Goldstone modes. One might be concerned that interactions with the gapless Majorana fermions could induce possibly long-range interactions between the Goldstone modes, which would invalidate the procedure of integrating out these Goldstone modes, or at least renormalize their properties such as stiffness and velocity, which would complicate the choice of parameters in the Goldstone mode Lagrangian. These effects, however, cannot happen because the stiffness and velocity are properties of the (3+1)D bulk while the Majorana fermions live in 2+1 dimensions. Deep in the ordered (superfluid) phase, the Goldstone modes interact weakly and are described by free massless bosons in 3+1 dimensions. The procedure of integrating out these free massless modes can thus be carried out exactly. A similar situation arises in the study of gauge field fluctuations in 3D topological Mott insulators\cite{witczak-krempa2010} and phonons in 3D topological insulators.\cite{habe2014}

In the imaginary-time formalism, the action for the bosonic Goldstone fields $\theta_{\alpha}$ is
\begin{align}
S_B  &= \int_0^{\beta} d\tau \int d^3 \mb R \, \mc L_B + \int_0^{\beta} d\tau H_{\text{coupling}},
\end{align}
where
\begin{subequations}
\begin{align}
\mc L_B &= \mc L_{0}(\partial_{\tau}\theta_{\alpha}) + \mc L_{\text{bend}}(\partial_i\theta_{\alpha}) + \mc L_{\text{dipole}}(\theta_x),\label{boson_lagrangian} \\
\mc L_0 &= \frac{1}{2} K_0 (\partial_{\tau} \theta_j)^2, \\
\mc L_{\text{bend}}& = \frac{1}{4}(K_T + K_L) (\partial_j \theta_k)^2 + \frac{1}{4}(K_T - K_L) \partial_j \theta_k \partial_k \theta_j , \\
\mc L_{\text{dipole}} & =   \frac{1}{2}g_D\, \theta_x^2,
\end{align}
\end{subequations}
where $\beta$ is the inverse temperature. The Lagrangian density $\mc L_B$ is composed of three distinct contributions. $\mc L_0$ contains the conjugate momenta for $\theta_{\alpha}$, $\mc  L_{\text{bend}}$ is the energy cost for having gradients of the bosonic fields,\cite{DeGennes1974} and $\mc L_{\text{dipole}}$ is the energy cost due to the nuclear dipole interaction between $\,^3$He quasiparticles.\cite{brinkman1974b} $\theta_x$ can be understood as the deviation of $\theta$ in Eq.~(\ref{RotMatrix}) from its equilibrium value given by the Leggett angle $\theta_L$. $K_T$ and $K_L = 3 K_T$ are the transverse and longitudinal stiffness, respectively, where $K_T=(2/5)N_F \xi_0^2 \Delta_0^2$ in weak-coupling theory\cite{wolfle1974} and $N_F = m k_F/\pi^2$ is the density of states at the Fermi energy. The constant $g_D$ is given by
\begin{align}
g_D =  3\lambda_D N_F \Delta_0^2,
\end{align}
where $\lambda_D\approx  5\times 10^{-7}$ is an approximately pressure-independent dimensionless constant.\cite{VollhardtWolfle} The dipole interaction produces a small energy gap $\propto\sqrt{g_D/K_0}$ for the $\theta_x$ fluctuations that acts as an infrared cutoff. The bending energy can be written in a more physical way,\cite{DeGennes1974}
\begin{equation}
\mc L_{\text{bend}}
= \frac{1}{4} K_L(\boldsymbol \nabla\cdot \mb R_{\mu})^2
+ \frac{1}{4} K_T (\boldsymbol \nabla\times \mb R_{\mu})^2,
\end{equation}
where we represented the rotation matrix $R_{\mu j}$ in Eq.~(\ref{small_order}) as a vector $\mb R_{\mu}$ for each $\mu$.

It remains to specify boundary conditions on the sample surface for the Goldstone field $\theta_x$ to be integrated out. (The $\theta_y$ and $\theta_z$ fields can be formally integrated out as well, but do not generate effective surface interactions for the Majorana fermions since they do not couple to the latter.) The spin supercurrent density\cite{brinkman1978} is defined in terms of the Lagrangian Eq.~(\ref{boson_lagrangian}) by
\begin{equation}
j_{\text{sp},i}^{\alpha} \propto \frac{\partial \mc L_B}{\partial (\partial_i \theta_{\alpha})},
\end{equation}
and corresponds to the supercurrent of the $\alpha$ component of spin flowing along direction $i$. The derivative with respect to $\partial_x \theta_x$ gives a term proportional to $\partial_x \theta_x$. Assuming that no spin supercurrent can escape from the $^3$He surface into the surrounding vacuum by flowing perpendicular to this surface, we impose the Neumann boundary condition $\partial_x\theta_x(x,y,z) |_{x=0}= 0$. With this boundary condition, $\theta_x(\mb Q)$ is even in $Q_x$ and thus couples in Eq.~(\ref{HcouplingThetax}) only to the part of $\rho(\b{Q})$ that is even in $Q_x$. After dropping the part that is odd in $Q_x$, the Majorana bilinear $\rho(\b{Q})$ in Eq.~(\ref{rho}) becomes
\begin{align}
&\rho(\b{Q})
=  f(Q_x) \rho(\mb Q_{\parallel})
,\quad
f(Q_x) = \frac{1}{1+ (Q_x/\kappa)^2}
\label{rho_even}.
\end{align}

The procedure of integrating out $\theta_x$ is best carried out in a frequency-momentum representation. We define the Fourier transform of $\theta_x(\tau,\mb R)$ by
\begin{align}
\theta_x (\nu_n, \mb Q)
 &=\int_0^{\beta}d\tau \int d^3 \mb R\, e^{-i\nu_n \tau }e^{i\mb Q \cdot \mb R} \theta_x(\tau,\mb R) ,\nonumber \\
\theta_x (\tau,\mb R)
 &=  \frac{1}{\beta V} \sum_{\nu_n, \mb Q} e^{i\nu_n \tau }e^{-i\mb Q \cdot \mb R} \theta_x(\nu_n,\mb Q),
\end{align}
where $\nu_n = 2\pi n/\beta, \, n\in \mathbb Z$ is a bosonic Matsubara frequency. Likewise, we define
\begin{align}
\rho(\nu_n,\mb Q)& =
\int_0^{\beta} d\tau \int  d^3\mb R\,  e^{-i\nu_n\tau}e^{i\mb Q\cdot \mb R}\rho(\tau,\mb R),\nonumber\\
\rho(\tau,\mb R)& = \frac{1}{\beta  V} \sum_{\nu_n,\mb Q} e^{i\nu_n\tau} e^{-i\mb Q\cdot \mb R}\rho(\nu_n,\mb Q),
\end{align}
for the Majorana bilinear. Performing the Gaussian path integral over $\theta_x$,
\begin{align}
\int\mathcal{D}\theta_x\,e^{-S_B[\theta_x,\rho]}\propto e^{-S_I[\rho]},
\end{align}
we obtain the action $S_I$ for an effective interaction between the surface Majorana fermions,
 \begin{equation}
 S_{\text{I}} = - \frac{\Delta_0^2}{2\beta V} \sum_{Q } \rho(-Q) G_{xx} (Q) \rho(Q),
 \end{equation}
 where we denoted the bulk (3+1)D frequency-momentum vector by $Q = (\nu_n, \mb Q)$, and $G_{xx}(Q)$ is the Green's function for $\theta_x$,
 \begin{align}
 G_{xx} (Q)^{-1} &=  K_T Q_x^2 + \bar G(Q_{\parallel})^{-1}, \nonumber\\
 \bar G(Q_{\parallel})^{-1}& =  K_S \mb Q_{\parallel}^2+ K_0 \nu_n^2 + g_D  ,
 \end{align}
where $K_S = (K_L+K_T)/2$, and we denoted the surface (2+1)D frequency-momentum vector by $Q_{\parallel} = (\nu_n, \mb Q_{\parallel})$. Since $\rho(\tau,\mb Q_{\parallel})$ does not depend on $Q_x$, the summation over $Q_x$ can be performed. If the thickness of the helium sample $L_\perp$ is large enough (we will comment shortly on the validity of this assumption), we can approximate the sum by an integral. We obtain
\begin{equation}\label{SI}
 S_{\text{I}} = - \frac{\Delta_0^2}{2 \beta V} \sum_{Q_{\parallel} } \rho(-Q_{\parallel}) G_{\parallel} (Q_{\parallel}) \rho(Q_{\parallel}),
 \end{equation}
where
\begin{align}\label{Gparallel}
 G_{\parallel}( Q_{\parallel})
& \equiv \sum_{Q_x} f(Q_x)^2 G_{xx}(Q) \nonumber \\
& = \bar G(Q_{\parallel})  \frac{ \kappa  L_{\perp}}{8}\frac{1+ 2\kappa(  K_T \bar G(Q_{\parallel}))^{1/2}}{\left[1 + \kappa  (K_T \bar G(Q_{\parallel}))^{1/2}\right]^2}.
\end{align}
The term in Eq.~(\ref{SI}) that is most relevant in the renormalization group sense is obtained by setting $Q_\parallel=0$ in the propagator $G_\parallel(Q_\parallel)$,
\begin{align}
 G_{\parallel}(0)= g_0' L_{\perp},\hspace{5mm} g_0' =\frac{\kappa  }{ g_D } \frac{1}{8} \frac{1+ 2\kappa\xi_D }{(1 + \kappa  \xi_D )^2},
\end{align}
where we defined the length scale $\xi_D = \sqrt{K_T/g_D}$ that may be called a ``dipole coherence length".\cite{VollhardtWolfle} It is the finite correlation length associated with the gapped mode $\theta_x$.

Approximating the sum over $Q_x$ by an integral as we have done in Eq.~(\ref{Gparallel}) is valid if $L_\perp$ is much larger than all other length scales in the problem. Because we have set $Q_\parallel=0$, the only other length scales are the superfluid coherence length $\xi_0\sim\kappa^{-1}$ and the dipole coherence length $\xi_D$. Using the weak-coupling expressions and a critical temperature of $T_c\approx 2.6$~mK at melting pressure,\cite{VollhardtWolfle} one obtains $\xi_0\approx 12$~nm and $\xi_D\approx 6.2$~$\mu$m. Since $\xi_D\gg\xi_0$ already we only require that $L_\perp\gg\xi_D$, i.e., the thickness of the helium sample should be much larger than a few microns. This is certainly the case in some experiments (e.g., Ref.~\onlinecite{murakawa2009}). In the case of thin films of $^3$He with $L_\perp$ on the order of a few microns (see, e.g., Ref.~\onlinecite{tsutsumi2011} and references therein), one should perform the sum over discrete values of $Q_x$ and also consider interactions induced by the Goldstone mode $\theta_x$ between Majorana fermions on opposite surfaces. For thicknesses comparable to the coherence length $\xi_0$, which is also the penetration depth of the surface states into the bulk, the surface states on opposite surfaces can trivially hybridize and open a gap without breaking any symmetries.

Assuming $L_\perp\gg\xi_D$, the effective interaction Hamiltonian $H_\textrm{I}$ corresponding to $S_\textrm{I}$ is
\begin{align}\label{HintSurf}
 H_{\text{I}}  = - \frac{g_0}{2}\sum_{\mb Q_{\parallel} } \rho(-\mb Q_{\parallel})  \rho(\mb Q_{\parallel}),
 \end{align}
where
\begin{align}
g_0& = g_0'  \frac{\Delta_0^2 }{ L_{\parallel}^2}
  = \frac{\kappa}{ N_F L_{\parallel}^2} \frac{1}{24 \lambda_D}  \frac{1+ 2\kappa\xi_D}{(1 + \kappa \xi_D)^2}. \label{g_0}
\end{align}
Equation (\ref{HintSurf}), the main result of our work, shows that the bulk Goldstone modes of $^3$He can induce effective short-range interactions between the surface Majorana fermions. Using the numerical values of parameters quoted above, we have $\kappa\xi_D\sim\xi_D/\xi_0\gg 1$ and the coupling constant $g_0$ (with units of energy) simplifies to
\begin{align}\label{g0simplified}
g_0\approx\frac{\Delta_0^2}{4L_\parallel^2\xi_Dg_D}.
\end{align}

What is the effect of these interactions on the physical properties of the surface Majorana fermions? The short-range interaction (\ref{HintSurf}) is perturbatively irrelevant at the free Majorana fermion fixed point Eq.~(\ref{H0Majorana}), hence the surface states are stable against this interaction if $g_0$ is sufficiently small.\cite{schnyder2008} The Majorana surface states may however become unstable if $g_0$ is sufficiently large. One exotic possibility is that the surface may undergo a transition to a state with non-Abelian topological order,\cite{fidkowski2013,metlitski2014} which preserves the symmetries of the free Majorana fermion state. The other, more conventional possibility is that the surface may spontaneously break some symmetries of the free Majorana fermion state. In the remainder of the paper we will focus on this possibility. For simplicity we will drop the subscript $\parallel$ on 2D spatial momenta, e.g., $\mb k_{\parallel}\rightarrow \mb k$ and $\mb Q_{\parallel}\rightarrow \mb Q$, given that the bulk has been integrated away and we are working with an effective 2D theory.

\section{Broken-symmetry states}
\label{sec:BrokenSymmetry}

In this section we study possible broken-symmetry states of surface Majorana fermions by using zero-temperature mean-field theory. We begin by identifying the possible order parameters. Restricting ourselves to translationally invariant Majorana fermion bilinears up to linear order in momentum, there are only three possibilities: a $\c{T}$-breaking mass order parameter $\c{M}$, a vector order parameter $\b{\c{V}}$ that breaks $\c{T}$ and rotational symmetry, and a nematic order parameter $\c{Q}_{ab}$ that breaks rotational symmetry. We find that an interaction of the form (\ref{HintSurf}) can lead to a first-order transition to a $\c{T}$-breaking state with $\langle\c{M}\rangle\neq 0$.

\subsection{Order parameters}

The simplest types of order parameters that can be constructed from Majorana fermions are fermion bilinears. We restrict ourselves to translationally-invariant order parameters,
\begin{align}
\mathcal{O}=\sum_\b{k}\gamma^T_{-\b{k}}
O(\b{k})\gamma_{\b{k}},
\end{align}
where $O(\b{k})$ is a Hermitian $2\times 2$ matrix that obeys $O(\b{k})=-O(-\b{k})^T$ due to Fermi statistics. For simplicity we will only consider terms of zeroth or first order in $\b{k}$.

Order parameters can be organized into representations of the symmetry group of the Hamiltonian
\begin{align}\label{HamiltonianH0HI}
H&=H_0+H_\text{I}\nonumber\\
&=\frac{v}{2}\sum_{\mb k}\gamma^{T}_{-\mb k }(\mb k\cdot\tilde{\bs \sigma})\gamma_{\mb k }
- \frac{g_0}{2}\sum_\b{Q}\rho(-\mb Q)  \rho(\mb Q),
\end{align}
where $v\equiv\Delta_0/k_F$ is the Majorana fermion velocity, hence one first needs to determine the symmetries of $H$. Besides translation invariance, $H$ is invariant under time-reversal symmetry defined by
\begin{equation}
\mathcal{T}\gamma_{\mb k\sigma}\mathcal{T}^{-1} = i\sigma^y_{\sigma\sigma'} \gamma_{-\mb k\sigma'},
\end{equation}
and under $SO(2)$ rotations by an angle $\theta\in[0,2\pi)$ about the surface normal $\hat{\b{x}}$, defined by
\begin{align}
\mathcal{R}(\theta)\gamma_{\b{k}\sigma}\mathcal{R}(\theta)^{-1}=R(\theta/2)_{\sigma\sigma'}\gamma_{R(-\theta)\b{k}\sigma'},
\end{align}
where the $2\times 2$ orthogonal representation matrix $R(\theta)$ is
\begin{align}
R(\theta)=\left(\begin{array}{cc}
\cos\theta & -\sin\theta \\
\sin\theta & \cos\theta
\end{array}\right).
\end{align}
The same representation matrix that acts on the spatial label $\b{k}$ also acts on the spinor label $\sigma$, but with half the angle. This is simply the statement that the Majorana field $\gamma$ forms a spinor representation of $SO(2)$, i.e., a representation of the double cover Spin(2). Because $R(\theta)$ is real, the reality condition Eq.~(\ref{reality}) is preserved under rotations.

We now enumerate the possible order parameters. To zeroth order in $\b{k}$ we can only have $O(\b{k})\propto\sigma^y$,
\begin{equation}
\mathcal{M} = \frac{1}{2}\sum_{\mb k}  \gamma^T_{-\b{k}} \sigma^y \gamma_{\mb k}.
\end{equation}
The Majorana mass term\cite{qi2009} $\mathcal{M}$ is odd under $\mathcal{T}$ but is invariant under $SO(2)$ rotations, since $R(\theta/2)=e^{-i\theta\sigma^y/2}$ commutes with $\sigma^y$. To linear order in $\b{k}$, we have the six possibilities $O(\b{k})\in\{k_y,k_z,k_y\sigma^x,k_z\sigma^x,k_y\sigma^z,
k_z\sigma^z\}$. By taking appropriate linear combinations, these six order parameters can be organized according to their transformation properties under $SO(2)$ into two scalars, one vector, and one symmetric traceless tensor. The two scalars are
\begin{align}
\sum_\b{k}\gamma_{-\b{k}}^T k_a\tilde{\sigma}^a\gamma_\b{k}
\propto H_0,
\hspace{5mm}
\sum_\b{k}\gamma_{-\b{k}}^T\epsilon_{ab}k_a\tilde{\sigma}^b\gamma_\b{k}
\propto\rho(\b{0}),
\end{align}
with $a,b=1,2$. These terms do not break $\mathcal{T}$ either, and a nonzero expectation value for them only leads to a finite renormalization of the surface state velocity (accompanied by a rotation of the spatial coordinate system). We can thus ignore them. The vector order parameter
\begin{align}\label{VectorOP}
\boldsymbol{\mathcal{V}}=\frac{1}{2k_F}\sum_\b{k}\gamma_{-\b{k}}^T\b{k}\gamma_{\b{k}},
\end{align}
transforms under rotations as $\mathcal{R}(\theta)\mathcal{V}_a\mathcal{R}(\theta)^{-1}=R(\theta)_{aa'}\mathcal{V}_{a'}$ and is odd under $\mathcal{T}$. From the point of view of symmetries, it can be interpreted as an in-plane ferromagnetic order parameter. Finally, the symmetric traceless tensor order parameter
\begin{align}
\mathcal{Q}_{ab}=
\frac{1}{2k_F}\sum_\b{k}\gamma_{-\b{k}}^T(k_a\tilde{\sigma}^b
+k_b\tilde{\sigma}^a-\delta_{ab}\b{k}\cdot\tilde{\bsigma})\gamma_{\b{k}},
\end{align}
transforms under rotations as $\mathcal{R}(\theta)\mathcal{Q}_{ab}\mathcal{R}(\theta)^{-1}=R(\theta)_{aa'}R(\theta)_{bb'}\mathcal{Q}_{a'b'}$ and is even under $\mathcal{T}$. It is a nematic order parameter\cite{DeGennes} with two independent components $\mathcal{Q}_{11}=-\mathcal{Q}_{22}$, $\mathcal{Q}_{12}=\mathcal{Q}_{21}$ forming a headless vector that is invariant under rotations by $\pi$,
\begin{align}
\mathcal{R}(\pi)\left(\begin{array}{c}
\mathcal{Q}_{11} \\ \mathcal{Q}_{12}\end{array}\right)\mathcal{R}(\pi)^{-1}
=R(2\pi)\left(\begin{array}{c}
\mathcal{Q}_{11} \\ \mathcal{Q}_{12}\end{array}\right)=\left(\begin{array}{c}
\mathcal{Q}_{11} \\ \mathcal{Q}_{12}\end{array}\right).
\end{align}

\subsection{Mean-field theory}

Zero-temperature mean-field theory is based on the variational principle of quantum mechanics $E_0\leq E_\textrm{MF}(\lambda)$ where $E_0$ is the energy of the true ground state and $E_\textrm{MF}(\lambda)=\langle\Phi_0(\lambda)|H|\Phi_0(\lambda)\rangle$ is the expectation value of the full Hamiltonian $H$ in a family of trial ground states $|\Phi_0(\lambda)\rangle$ parameterized by a variational parameter $\lambda$. The optimal variational ground state is determined by minimizing $E_\mathrm{MF}(\lambda)$ with respect to $\lambda$, i.e., finding the solutions of $\partial_\lambda E_\textrm{MF}(\lambda)=0$. The trial states $|\Phi_0(\lambda)\rangle$ can be constructed as the ground states of a family of trial Hamiltonians $H_\textrm{MF}(\lambda)$. Applied to our problem, for each order parameter $\mathcal{O}$ in turn we define the trial Hamiltonian as
\begin{align}
H_\textrm{MF}(\lambda)=H_0+\lambda\mathcal{O},
\end{align}
which is quadratic in the Majorana fermions $\gamma$, hence can be solved exactly for $|\Phi_0(\lambda)\rangle$. The variational parameter $\lambda$ is the Legendre transform of the operator $\mathcal{O}$, and is proportional to $\langle\mathcal{O}\rangle$ --- hence it is often also called the order parameter. In what follows we use Latin letters $M$, $\b{V}$, $Q_{ab}$ to denote the corresponding variational parameters $M\sim\langle\mathcal{M}\rangle$, $\b{V}\sim\langle\boldsymbol{\mathcal{V}}\rangle$, $Q_{ab}\sim\langle\mathcal{Q}_{ab}\rangle$. In this section we only outline the main steps of the mean-field calculations; technical details can be found in Appendix~\ref{sec:appendix}.

In principle, one should consider all order parameters simultaneously,
\begin{align}
H_\textrm{MF}(M,V,Q_{ab})=H_0+M\c{M}+\b{V}\cdot\b{\c{V}}+Q_{ab}\c{Q}_{ab},
\end{align}
and minimize $E_\textrm{MF}(M,V,Q_{ab})$ with respect to the 5D parameter space $\{M,V,Q_{ab}\}$. Here we will consider the simpler approach of studying each order parameter in turn. Our conclusion will be that the only relevant instability is the $\c{T}$-breaking mass instability; thus, the issue of phase coexistence is irrelevant to our discussion.

\begin{figure}
\includegraphics[width=1\linewidth]{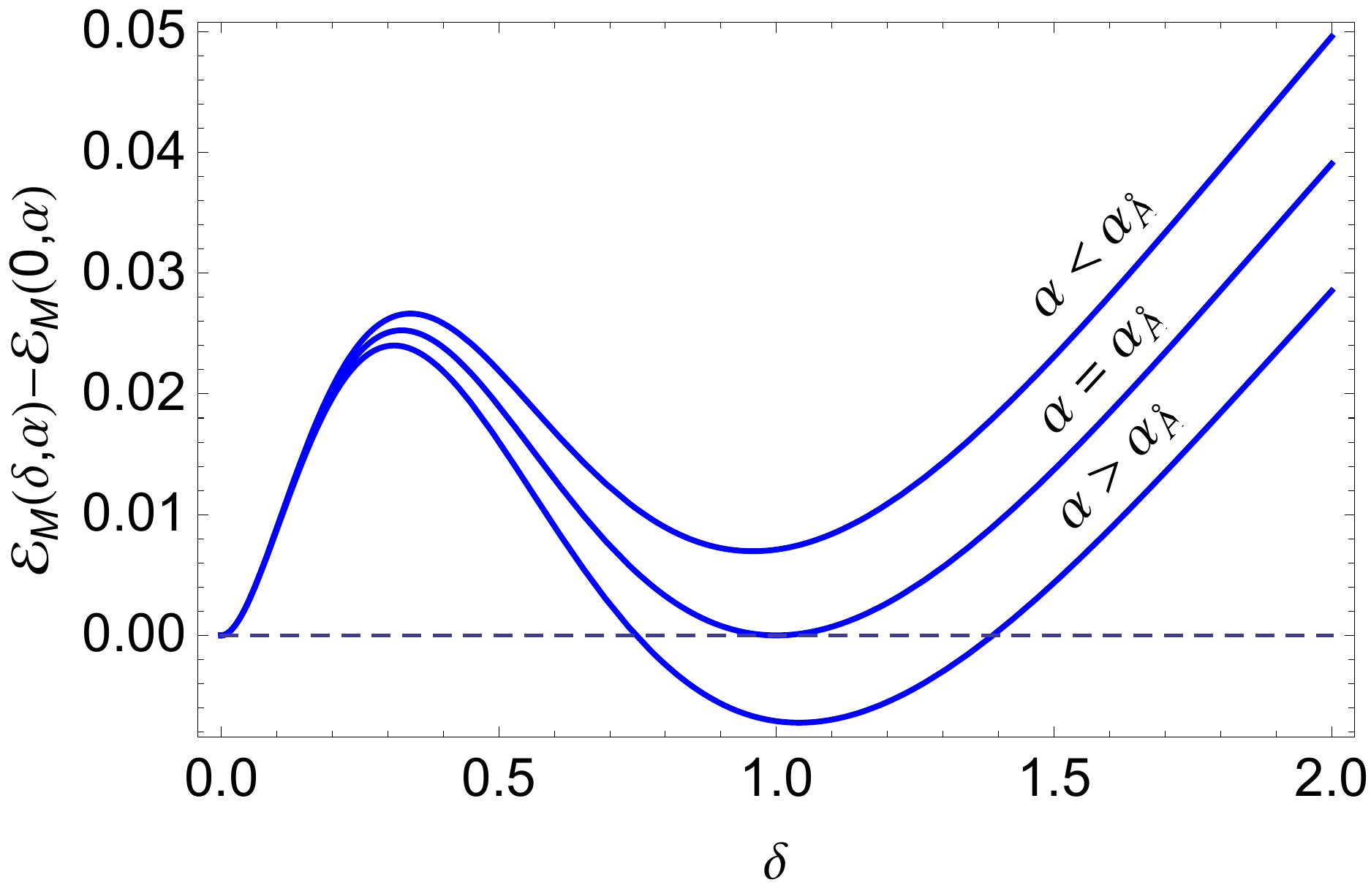}
\caption{Dimensionless variational energy (\ref{dimless_em}) as a function of the dimensionless $\c{T}$-breaking mass $\delta=M/v\Lambda$ and the dimensionless coupling constant $\alpha= gL_\parallel^2\Lambda^3/48\pi v$. There is a first-order transition at $\alpha=\alpha_c=(1+\sqrt{2})^2$.}
\label{fig:mass}
\end{figure}
To investigate the instability towards spontaneously generating a Majorana mass, we consider the mean-field Hamiltonian
\begin{align}
H_\textrm{MF}(M)=H_0+M\mathcal{M},
\end{align}
for which the variational energy $E_\textrm{MF}(M)$ is given in Eq.~(\ref{TotalEM}). Defining $g\equiv g_0/k_F^2$ where $g_0$ is the coupling constant in the surface state Hamiltonian (\ref{HamiltonianH0HI}), we find a first-order transition (Fig.~\ref{fig:mass}) at a critical value of $g$ given by
\begin{align}\label{gCriticalMass}
g_c=\frac{48(1+\sqrt{2})^2\pi v}{L_\parallel^2\Lambda^3},
\end{align}
below which $\langle\c{M}\rangle=0$ and above which $\langle\c{M}\rangle\neq 0$.

For the vector order parameter, we consider the mean-field Hamiltonian
\begin{align}\label{HMF_FM}
H_\textrm{MF}(\b{V})=H_0+\b{V}\cdot\b{\c{V}}.
\end{align}
The term $\b{V}\cdot\b{\c{V}}$ is equivalent to the Doppler shift induced on the surface by a bulk superflow with velocity $\b{v}_s=\b{V}/k_F$.\cite{wu2013} To see this explicitly, suppose that the fluid is flowing with the superfluid velocity $\bm v_s = (v_s^y,v_s^z)$ with respect to the wall. The BdG Hamiltonian in the rest frame $H_{\text{BdG}}'$ is obtained from a Galilean transformation $\epsilon_{\bm K} \ra \epsilon_{\bm K+ m \bm v_s}$ on Eq.~(\ref{bulk}) where $\bm K = (k_x,\bm k)$ denotes the 3D momentum,
\begin{align}\label{DopplerShift}
  H'_{\text{BdG}} = H_{\text{BdG}} + \frac{1}{2} \sum_{\bm K } (\bm v_s\cdot \bm k) \Psi_{\bm K}^{\dagger}\Psi_{\bm K}.
\end{align}
The $\bm v_s$-dependent term does not affect the spinor structure of the Majorana fermion operator, and we may continue to use the approximate form of the field operator Eq.~(\ref{field_op_exp}). The $\bm v_s$-dependent term then reduces to $\bm V \cdot \bm {\mc V}$ with $\bm V = k_F\bm v_s$. Therefore, a nonzero vector order parameter $\b{V}$ must be accompanied by a bulk phase gradient and does not correspond to an instability occurring only on the surface. We will discard it in the remainder of our analysis.

\begin{figure}
\includegraphics[width=1\linewidth]{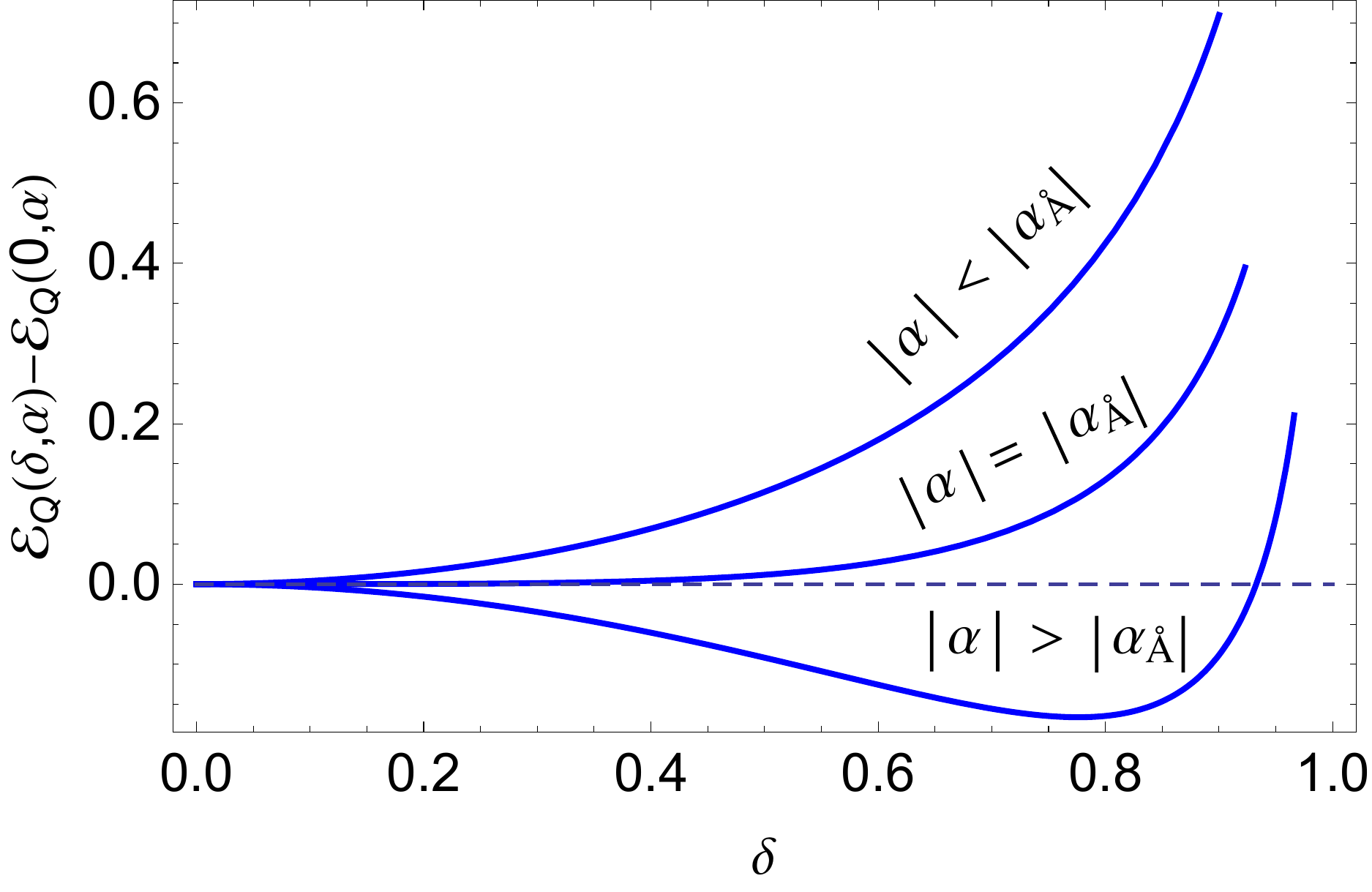}
\caption{Dimensionless variational energy (\ref{dimless_eq}) as a function of the dimensionless nematic order parameter $\delta=Q/vk_F$ and the dimensionless coupling constant $\alpha= gL_\parallel^2\Lambda^3/96\pi^2 v$. There is a continuous transition at $\alpha=\alpha_c=-\frac{3}{2}$.}
\label{fig:nematic}
\end{figure}
Finally, for nematic order we consider the mean-field Hamiltonian
\begin{align}
H_\textrm{MF}(Q_{ab})=H_0+Q_{ab}\c{Q}_{ab},
\end{align}
for which the variational energy $E_\textrm{MF}(Q_{ab})$ is given in Eq.~(\ref{EMFQ}). We find a continuous transition (Fig.~\ref{fig:nematic}) at a critical value of $g$ given by
\begin{align}
g_c=-\frac{144\pi^2v}{L_\parallel^2\Lambda^3},
\end{align}
such that $\langle\c{Q}_{ab}\rangle\neq 0$ for $g<g_c$ and $\langle\c{Q}_{ab}\rangle=0$ for $g>g_c$. However, $g_c$ is negative while the coupling constant (\ref{g0simplified}) is positive. Therefore, according to this calculation the surface of $^3$He-$B$ is necessarily in the isotropic phase.

\section{Discussion and conclusions}

Our mean-field calculation predicts that the massless Majorana fermions at the surface of $^3$He-$B$ could spontaneously develop a $\c{T}$-breaking mass if the coupling constant for the effective surface interaction (\ref{HintSurf}) mediated by the bulk spin-orbit collective modes exceeds a certain critical value given by Eq.~(\ref{gCriticalMass}). How do those two couplings compare? Denoting by $g_\textrm{He}\equiv g_0/k_F^2$ the coupling constant in $^3$He-$B$ with $g_0$ given in Eq.~(\ref{g0simplified}),
\begin{align}
g_\textrm{He}=\frac{\Delta_0^2}{4L_\parallel^2\xi_Dg_Dk_F^2},
\end{align}
we have
\begin{align}\label{gHegc}
\frac{g_\textrm{He}}{g_c}=\frac{\pi^2}{12\eta\lambda_D}\frac{v_F/\Delta_0}{\xi_D}\left(\frac{\Lambda}{k_F}\right)^3
=13.46\left(\frac{\Lambda}{k_F}\right)^3,
\end{align}
where $\eta\equiv 48(1+\sqrt{2})^2\pi$, and we have used the expressions given in Sec.~\ref{sec:EffSurfInt} for $g_D$, $\lambda_D$, $N_F$, $\xi_0$, and $\xi_D$. The large momentum cutoff $\Lambda$ is of the same order of magnitude as $k_F$. Therefore, although one cannot admittedly rely on mean-field theory for accurate predictions of critical coupling constants, Eq.~(\ref{gHegc}) nonetheless suggests that $g_\textrm{He}$ and $g_c$ are of the same order of magnitude. This implies the possibility that the surface Majorana fermions in $^3$He-$B$ may be in the vicinity of a quantum phase transition to a $\c{T}$-breaking phase as a result of their coupling to bulk collective modes, or possibly already in a $\c{T}$-breaking phase.

How does this prediction compare to experiments on $^3$He-$B$? Surface Andreev bound states in this system have been studied by various means over the past ten years or so.\cite{okuda2012} Transverse acoustic impedance measurements\cite{aoki2005,saitoh2006,nagai2008,wada2008,
murakawa2009,murakawa2009b,murakawa2011} are consistent with the existence of surface states with energies within the bulk superfluid gap. Specific heat measurements\cite{choi2006} and transverse sound attenuation measurements\cite{davis2008} independently support this conclusion. More specifically, the growth of a low-frequency peak in the transverse acoustic impedance with increasing specularity of the $^3$He-$B$ surface was interpreted in Ref.~\onlinecite{murakawa2011} as a signature of the linear energy dependence of the density of states of massless surface Majorana fermions, based on qualitative agreement with theoretical calculations. In all these studies however, the presence of a small gap in the surface state dispersion would be hard to detect, especially because a mass gap for 2D Majorana fermions is not accompanied by sharp band-edge features in the density of states---unlike, for example, the bulk gap of a 3D $s$-wave superconductor which is flanked by Bardeen-Cooper-Schrieffer (BCS) coherence peaks. Finally, even if the Majorana surface states are in fact in the gapless $\c{T}$-preserving phase, our work suggests that they may be strongly correlated and not adequately described as free Majorana fermions.

Our theory predicts a first-order transition, with a mass gap $M$ jumping from zero to a value equal to the bulk superfluid gap $\Delta_0$ at the transition (Fig.~\ref{fig:mass}). This would obviously contradict the experimental observations mentioned above of a nonzero density of states within the bulk gap, but is most likely an artefact of the mean-field approximation. Fluctuations are likely to reduce the jump in the order parameter, or could even make the transition continuous. If the latter happens, recent work\cite{ScottThomas,sonoda2011,grover2012,grover2014} has shown that this transition should exhibit an emergent $\c{N}=1$ supersymmetry (SUSY). Ref.~\onlinecite{grover2014} outlines an interesting proposal to induce a $\c{T}$-breaking transition on the surface of $^3$He-$B$ by applying a magnetic field perpendicular to the surface.\cite{mizushima2012} Our work suggests that $^3$He-$B$ may already be close to a $\c{T}$-breaking transition due to the coupling between surface Majorana fermions and bulk spin-orbit collective modes. This would suggest the alternate scenario of reaching such a transition by tuning bulk parameters, such as pressure, to vary the coupling constant $g_\textrm{He}$ in Eq.~(\ref{gHegc}) without breaking $\c{T}$ explicitly. In either scenario, one would need experimental probes able to detect the breaking of $\c{T}$ on the surface of $^3$He-$B$, such as perhaps the Magnus force technique used in Ref.~\onlinecite{ikegami2013}. We hope that our work, as well as the tantalizing prospect of discovering SUSY in a condensed matter system, will stimulate further experimental studies of surface states in $^3$He-$B$.

\acknowledgements

We thank J. P. Davis and A.-M. S. Tremblay for illuminating discussions.  This work was supported in part by the Simons Foundation and the Natural Sciences and Engineering Research Council (NSERC) of Canada (JM), the Institute for Basic Science (IBS) of Korea through the Young Scientist grant (SBC), and also supported  in part by the Department of Energy, Office of Basic Energy Sciences through grant No. DE-SC0002140 (YJP).


\appendix
\section{Details of the mean-field calculation}
\label{sec:appendix}

\subsection{Mass instability}
\label{sec:AppMass}

The mean-field Hamiltonian is
\begin{align}
  H_\textrm{MF}(M)
  & = H_0 + M \sum_{\bm k} \half \gamma_{-\bm k}^T \sigma^y \gamma_{\bm k} \nn\\
 & =\half \sum_{\bm k} \gamma_{-\bm k}^T
 \begin{pmatrix}
   v k_y & v k_z - i M \\
   v k_z + i M & - v k_y
 \end{pmatrix}
 \gamma_{\bm k},
\end{align}
where $M$ is a single variational parameter. The spectrum is $E_M(\bm k)  =  \sqrt{v^2 {\bm k}^2 + M^2}$. The Hamiltonian matrix has the structure
\begin{equation}
  \begin{pmatrix}
    \cos 2 \theta & e^{- i \varphi}\sin 2\theta \\
    e^{i \varphi}\sin 2\theta & -\cos 2\theta
  \end{pmatrix},
\end{equation}
with eigenvalues $\pm 1$, and eigenvectors
\begin{align}
|+\rangle &=
\begin{pmatrix}
  \cos \theta \\ e^{i\varphi}\sin \theta
\end{pmatrix}
=
\begin{pmatrix}
  u \\v
\end{pmatrix}
,\\
|-\rangle &=
\begin{pmatrix}
  e^{-i\varphi}\sin \theta\\ -\cos \theta
\end{pmatrix}
=
\begin{pmatrix}
  v^* \\ -u
\end{pmatrix}.
\end{align}
We have the identifications
\begin{align*}
v k_y & = E_M(\bm k) \cos 2\theta_{\bm k}
 = E_M(\bm k) ( u_{\bm k}^2-|v_{\bm k}|^2 ), \\
  v k_z +i M &= E_M(\bm k) e^{i\varphi_{\bm k}}\sin 2\theta_{\bm k}
  = E_M(\bm k)\,( 2u_{\bm k} v_{\bm k}),
\end{align*}
where we define $u_{\bm k} = \cos \theta_{\bm k}$ and $v_{\bm k} = e^{i\varphi_{\bm k}}\sin \theta_{\bm k}$ with
\begin{align}
\tan  \varphi_{\bm k} = \frac{M}{v k_z}, \quad
\cos 2\theta_{\bm k} = \frac{v k_y}{E_M(\bm k)}.
\end{align}
We also define the Hermitian and unitary matrix
\begin{equation}
  U(\bm k) =
  \begin{pmatrix}
    u_{\bm k} & v^*_{\bm k}\\
    v_{\bm k} & - u_{\bm k}
  \end{pmatrix}
  =U^{\dagger}(\bm k)=U^{-1}(\bm k),
\end{equation}
in terms of which the mean-field Hamiltonian becomes
\begin{align}
  H_{\MF}(M)
  &=
  \half \sum_{\bm k} \gamma_{-\bm k}^T U^{\dagger}(\bm k)
  \begin{pmatrix}
    E_M(\bm k) & 0 \\
    0& - E_M(\bm k)
  \end{pmatrix}
  U(\bm k) \gamma_{\bm k} \nn\\
  &= \half \sum_{\bm k} \eta_{\bm  k}^{\dagger} \begin{pmatrix}
    E_M(\bm k) & 0 \\
    0& - E_M(\bm k)
  \end{pmatrix} \eta_{\bm k} \nn\\
  &= \half \sum_{\bm k}\left( E_M(\bm k) \eta_{\bm k\ua}^{\dagger} \eta_{\bm k\ua}-E_M(\bm k) \eta_{\bm k\da}^{\dagger} \eta_{\bm k\da}\right),
\end{align}
where in the last line $\ua,\da$ do not denote spin but a band index. We define the eigenoperators
\begin{align}\label{etakdef}
  \eta_{\bm k} &
  = \begin{pmatrix} \eta_{\bm k\ua} \\ \eta_{\bm k\da} \end{pmatrix}
  = U(\bm k)  \gamma_{\bm k},
\end{align}
which satisfy the canonical anticommutation relations
\begin{equation}
 \{ \eta_{\bm k\alpha} ,\eta_{\bm k'\alpha'}^{\dagger}\} = \delta_{\bm k+\bm k',0} \delta_{\alpha\alpha'},
 \quad  \{ \eta_{\bm k\alpha} ,\eta_{\bm k'\alpha'}\} = 0.
\end{equation}
The Majorana fermion operators are given in terms of the $\eta_\b{k}$ as
\begin{align}
\gamma_\b{k}=U^\dag(\b{k})\eta_\b{k}=U(\b{k})\eta_\b{k},\quad
\gamma_{-\b{k}}=U^*(\b{k})(\eta^\dag_\b{k})^T.
\end{align}
The variational ground state $|\Phi_0(M)\rangle$ of $H_{\MF}(M)$ is defined by $\eta_{\bm k\ua } |\Phi_0(M)\rangle =  \eta_{\bm k\da }^{\dagger} |\Phi_0(M)\rangle = 0$, for all $\b{k}$. The total variational energy $E_\textrm{MF}(M)$ is given by the sum of the noninteracting energy $\langle\Phi_0(M)|H_0|\Phi_0(M)\rangle$ and the interaction energy $\langle\Phi_0(M)|V|\Phi_0(M)\rangle$. The noninteracting variational energy is
\begin{align}\label{E0M}
&\langle \Phi_0(M)|  H_0|\Phi_0(M)\rangle \nn\\
&= \half v\sum_{\bm k}[U(\bm k)\bm k\cdot \tilde \bsigma U^{\dagger}(\bm k)]_{\da\da}  \langle\Phi_0(M)| \eta_{\bm k\da}^{\dagger}\eta_{\bm k\da}|\Phi_0(M)\rangle\nn\\
&=- \half \sum_{\bm k} \frac{v^2\bm k^2}{E_M(\bm k)}.
\end{align}
Denoting the $2\times 2$ matrix $\bm k\times \tilde{\bm \sigma}$ by $w(\bm k)$, the interaction is
\begin{equation}
  V = -\frac{g}{8} \sum_{\bm k \bm k'\bm q}
   \gamma_{-\bm k + \half \bm q}^T w(\bm k) \gamma_{\bm k + \half \bm q}
    \gamma_{-\bm k'- \half \bm q}^T w(\bm k') \gamma_{\bm k'-  \half \bm q},
\end{equation}
where we have defined $g=g_0/k_F^2$ for simplicity. The interaction variational energy is
\begin{align}
\langle\Phi_0(M)|V|\Phi_0(M)=-\frac{g}{8}\sum_{\b{k}\b{k}'\b{q}}&w_{\alpha\beta}(\b{k})w_{\alpha'\beta'}(\b{k}')\nn\\
&\times\Gamma_{\alpha\beta\alpha'\beta'}^{(M)}(\b{k},\b{k}',\b{q}),
\end{align}
where we define the four-point function
\begin{align}\label{4pointfunction}
&\Gamma_{\alpha\beta \alpha'\beta'}^{(M)}(\bm k,\bm k',\bm q)=  \nn\\
& \langle \Phi_0(M)| \gamma_{-\bm k + \half \bm q,\alpha} \gamma_{\bm k + \half \bm q,\beta}
    \gamma_{-\bm k'-\half \bm q,\alpha'} \gamma_{\bm k'-  \half \bm q,\beta'}|\Phi_0(M)\rangle.
\end{align}
Applying Wick's theorem to Eq.~(\ref{4pointfunction}) yields contributions in the three interaction channels $\b{q}=0$, $\b{k}-\b{k}'=0$, and $\b{k}+\b{k}'=0$. In the $\b{q}=0$ channel, we have
\begin{widetext}
\begin{align}\label{Channel1}
\Gamma_{\alpha\beta\alpha'\beta'}^{(M)}(\bm k,\bm k',0)&=\langle \Phi_0(M)| \gamma_{-\bm k,\alpha} \gamma_{\bm k,\beta}
    \gamma_{-\bm k',\alpha'} \gamma_{\bm k',\beta'}|\Phi_0(M)\rangle
    \nn\\
&= U_{\da\alpha}(\bm k)U_{\beta\da}(\bm k) U_{\da\alpha'}(\bm k')U_{\beta'\da}(\bm k')\langle \Phi_0(M)| \eta_{\bm k\da}^{\dagger}\eta_{\bm k\da}\eta_{\bm k'\da}^{\dagger}  \eta_{\bm k'\da}|\Phi_0(M)\rangle
= U_{\da\alpha}(\bm k)U_{\beta\da}(\bm k) U_{\da\alpha'}(\bm k')U_{\beta'\da}(\bm k').
\end{align}
In the $\bm k - \bm k'=0$ channel, we have
\begin{align}\label{Channel2}
&\Gamma_{\alpha\beta \alpha'\beta'}^{(M)}(\bm k,\bm k,\bm q)  =  \langle \Phi_0(M)| \gamma_{-(\bm k - \half \bm q),\alpha} \gamma_{\bm k + \half \bm q,\beta}
    \gamma_{-(\bm k+\half \bm q),\alpha'} \gamma_{\bm k-  \half \bm q,\beta'}|\Phi_0(M)\rangle\nn\\
&=U_{\da \alpha}(\bm k -\half \bm q) U_{\da\beta}(-\bm k -\half \bm q) U_{\alpha'\da }(-\bm k- \half \bm q)U_{\beta'\da}(\bm k- \half \bm q) \langle \Phi_0(M)| \eta_{\bm k - \half \bm q\da}^{\dagger} \eta_{\bm k-\half \bm q\da} \eta_{-(\bm k + \half \bm q)\da}^{\dagger}
\eta_{-(\bm k+\half \bm q)\da}|\Phi_0(M)\rangle\nn\\
&\quad +\delta_{\alpha'\beta'} \delta_{\bm q,0} U_{\da\alpha}(\bm k)U_{\beta\da}(\bm k) \langle \Phi_0(M)| \eta_{\bm k \da}^{\dagger} \eta_{\bm k \da}|\Phi_0(M)\rangle -\delta_{\beta\beta'} \delta_{\bm k,0}U_{\da \alpha}(-\half \bm q)U_{\alpha'\da}(-\half\bm q)\langle \Phi_0(M)| \eta_{-\half\bm q \da}^{\dagger} \eta_{-\half\bm q\da} |\Phi_0(M)\rangle\nn\\
&=U_{\da \alpha}(\bm k -\half \bm q) U_{\da\beta}(-\bm k -\half \bm q) U_{\alpha'\da }(-\bm k- \half \bm q)U_{\beta'\da}(\bm k- \half \bm q) +\delta_{\alpha'\beta'} \delta_{\bm q,0} U_{\da\alpha}(\bm k)U_{\beta\da}(\bm k)-\delta_{\beta\beta'} \delta_{\bm k,0}U_{\da \alpha}(-\half \bm q)U_{\alpha'\da}(-\half\bm q).
\end{align}
Finally, in the $\b{k}+\b{k}'=0$ channel we have
\begin{align}\label{Channel3}
&\Gamma_{\alpha\beta \alpha'\beta'}^{(M)}(\bm k,-\bm k,\bm q) =  \langle \Phi_0(M)| \gamma_{-(\bm k - \half \bm q),\alpha} \gamma_{\bm k + \half \bm q,\beta}
    \gamma_{\bm k-\half \bm q,\alpha'} \gamma_{-(\bm k+ \half \bm q),\beta'}|\Phi_0(M)\rangle\nn\\
&=  -U_{\da \alpha}(\bm k -\half \bm q)U_{\alpha'\da}(\bm k - \half \bm q) U_{\da \beta}(-\bm k -\half \bm q)U_{\beta'\da}(-\bm k -\half \bm q)\langle\Phi_0(M)| \eta_{\bm k - \half \bm q\da}^{\dagger}
    \eta_{\bm k-\half \bm q\da} \eta_{-(\bm k + \half \bm q)\da}^{\dagger} \eta_{-(\bm k+ \half \bm q)\da}|\Phi_0(M)\rangle\nn\\
    & \quad +\delta_{\beta\alpha'} \delta_{\bm k,0}U_{\da \alpha}(-\half \bm q)U_{\beta'\da}(-\half\bm q)\langle\Phi_0(M)| \eta_{-\half\bm q \da}^{\dagger} \eta_{-\half\bm q\da} |\Phi_0(M)\rangle\nn\\
    &=  -U_{\da \alpha}(\bm k -\half \bm q) U_{\da \beta}(-\bm k -\half \bm q)U_{\alpha'\da}(\bm k - \half \bm q)U_{\beta'\da}(-\bm k -\half \bm q)+ \delta_{\beta\alpha'} \delta_{\bm k,0}U_{\da \alpha}(-\half \bm q)U_{\beta'\da}(-\half\bm q).
\end{align}
\end{widetext}
Ignoring terms independent of the order parameter $M$, we obtain
\begin{align}\label{EVM}
\langle\Phi_0(M)|V|\Phi_0(M)\rangle=- \frac{g}{16}\left(\sum_{\bm p}E_M(\bm p)\right)\left(\sum_{\bm p} \frac{\bm p^2}{E_M(\bm p)}\right).
\end{align}
Converting momentum sums to integrals in the limit of large $L_\parallel$, we have
\begin{align}
\sum_\b{p}E_M(\b{p})&=\frac{L_\parallel^2}{2\pi}\int_0^\Lambda dp\,p\sqrt{v^2p^2+M^2}\nn\\
&=\frac{vL_\parallel^2\Lambda^3}{6\pi}\left[(1+\delta^2)^{3/2}-|\delta|^3\right],
\end{align}
and
\begin{align}
\sum_{\bm p} \frac{\bm p^2}{E_M(\bm p)}&=\frac{L_\parallel^2}{2\pi}\int_0^\Lambda\frac{dp\,p^3}{\sqrt{v^2p^2+M^2}}\nn\\
&=\frac{L_\parallel^2\Lambda^3}{6\pi v}\left[(1-2\delta^2)(1+\delta^2)^{1/2}+2|\delta|^3\right],
\end{align}
where we have introduced a large-momentum cutoff $\Lambda$ and we define the dimensionless order parameter $\delta\equiv M/v\Lambda$. Adding the noninteracting (\ref{E0M}) and interaction (\ref{EVM}) contributions, the total variational energy is given by
\begin{align}\label{TotalEM}
E_\textrm{MF}(M)=\frac{v L_\parallel^2\Lambda^3}{12\pi}\c{E}_M(\delta,\alpha),
\end{align}
where the dimensionless function $\c{E}_M(\delta,\alpha)$ is defined as
\begin{align}
  \c{E}_M(\delta,\alpha) =& -\left((1-2\delta^2)(1+ \delta^2)^{1/2}+2|\delta|^3\right)\nn\\
  &\times
 \left[1+\alpha\left( (1+ \delta^2)^{3/2}-|\delta|^3\right)\right], \label{dimless_em}
\end{align}
with $\alpha\equiv gL_\parallel^2\Lambda^3/48\pi v$ a dimensionless coupling constant.

Minimizing $E_\textrm{MF}(M)$ with respect to $M$ is equivalent to minimizing $\c{E}_M(\delta,\alpha)$ with respect to $\delta$. We find two local minima, one at $\delta=0$ corresponding to the disordered, $\c{T}$-invariant phase and one at $\delta\neq 0$ corresponding to the ordered, $\c{T}$-breaking phase. There is a first-order transition at a critical value of $\alpha$ given by $\alpha_c=(1+\sqrt{2})^2$ at which $\delta=\delta_c=1$. For $\alpha<\alpha_c$, $\delta=0$ is the lowest-energy solution while for $\alpha>\alpha_c$, $\delta\neq 0$ has lowest energy (Fig.~\ref{fig:mass}). This corresponds to a critical coupling constant
\begin{align}
g_c=\frac{48(1+\sqrt{2})^2\pi v}{L_\parallel^2\Lambda^3},
\end{align}
below which $M=0$ and above which $M\neq 0$.

\subsection{Nematic instability}
\label{sec:AppNematic}

The mean-field Hamiltonian is $H_\textrm{MF}(Q_{ab})=H_0+Q_{ab}\c{Q}_{ab}$. As in the case of the ferromagnetic instability, we can use rotational invariance to set $Q_{ab}=(Q_{11},Q_{12})=(Q,0)$ for the purposes of computing the variational energy. We have
\begin{align}
  H_\textrm{MF}(Q)
  & = H_0 + \frac{Q}{k_F} \sum_{\bm k} \half \gamma_{-\bm k}^T
(k_y\sigma^z-k_z\sigma^x) \gamma_{\bm k} \nn\\
 & =\half \sum_{\bm k} \gamma_{-\bm k}^T
 \begin{pmatrix}
   \left(v+\frac{Q}{k_F}\right) k_z & \left(v-\frac{Q}{k_F}\right) k_x  \\
   \left(v-\frac{Q}{k_F}\right) k_x  & - \left(v+\frac{Q}{k_F}\right) k_z
 \end{pmatrix}
 \gamma_{\bm k},
\end{align}
where $Q$ is a single variational parameter. The spectrum is $E_Q(\bm k)  =  v\sqrt{\alpha^2 k_y^2 + \beta^2 k_z^2}$ where we define $\alpha = 1 + \delta$, $\beta = 1 - \delta$ and $\delta\equiv Q/vk_F$.
The Hamiltonian matrix has the same structure and eigenvectors as Eq.~(\ref{HstructureFM}) and (\ref{HeigenvectorsFM}). We have the identifications
\begin{align*}
 \alpha v k_y & = E_Q(\bm k) \cos 2\theta_{\bm k}
 = E_Q(\bm k) ( u_{\bm k}^2-v_{\bm k}^2 ), \\
  \beta v k_z  &= E_Q(\bm k) \sin 2\theta_{\bm k}
  = E_Q(\bm k)\,( 2u_{\bm k} v_{\bm k}),
\end{align*}
or, solving for $\theta_{\bm k}$,
\begin{align*}
\tan 2\theta_{\bm k} &= \frac{\beta}{\alpha}\frac{k_z}{k_y}.
\end{align*}
The Hamiltonian is diagonalized by a Hermitian and unitary matrix $U(\b{k})$ defined as in Eq.~(\ref{UkFerromagnet}). We obtain
\begin{align}
  H_\textrm{MF}(Q)
  = \half \sum_{\bm k}\left( E_Q(\bm k) \eta_{\bm k \ua}^{\dagger} \eta_{\bm k \ua}-E_Q(\bm k) \eta_{\bm k \da}^{\dagger} \eta_{\bm k\da}\right),
  \end{align}
as before, with the definition (\ref{etakdef}).

In order to evaluate momentum integrals, it is convenient to introduce the change of variables $k_y = |\b{k}| \cos \varphi$ and $k_z = |\b{k}| \sin \varphi$, in terms of which the energy spectrum becomes
\begin{align}
  E_Q(\bm k)
  &= v |\b{k}| (1 + \delta^2 + 2\delta \cos 2\varphi)^{1/2}\nn\\
  &= v |\b{k}| (1+\delta)(1 -   \delta_2 \sin^2\varphi)^{1/2},
\end{align}
where we define $\delta_2\equiv 4\delta/(1+\delta)^2$. The noninteracting variational energy is
\begin{align}\label{E0Q}
&\langle \Phi_0(Q)|  H_0|\Phi_0(Q)\rangle \nn\\
&\quad= \half v\sum_{\bm k}[U(\bm k)\bm k\cdot \tilde{\b{\sigma}} U^{\dagger}(\bm k)]_{\da \da}  \langle \Phi_0(Q)| \eta_{\bm k\da}^{\dagger}\eta_{\bm k\da}|\Phi_0(Q)\rangle\nn\\
&\quad=- \half v\sum_{\bm k}\left( k_z (2u_{\bm k}v_{\bm k})+k_y(u_{\bm k}^2 - v_{\bm k}^2)\right)\nn\\
&\quad=- \half v^2\sum_{\bm k} \frac{\alpha k_y^2+\beta k_z^2}{E_Q(\bm k)}\nn\\
&\quad= -  \frac{v}{2(1+\delta)}\sum_{\bm k} \frac{1+\delta-2\delta\sin^2\varphi}{(1 -  \delta_2 \sin^2\varphi)^{1/2}}|\b{k}|\nn\\
&\quad=\frac{v}{\pi}\left(\frac{\delta^2}{1+\delta}K(\delta_2)-(1+\delta)E(\delta_2)\right)\sum_\b{k}|\b{k}|,
\end{align}
where $K$ and $E$ are the complete elliptic integrals of the first and second kind, respectively, defined as
\begin{align}
K(m)&=\int_0^{\pi/2}\frac{d\varphi}{(1-m\sin^2\varphi)^{1/2}},\label{EllipticK}\\
E(m)&=\int_0^{\pi/2}d\varphi\,(1-m\sin^2\varphi)^{1/2}.\label{EllipticE}
\end{align}

To compute the interaction variational energy, we define a four-point function as in Eq.~(\ref{4pointfunction}),
\begin{align}
&\Gamma_{\alpha\beta \alpha'\beta'}^{(Q)}(\bm k,\bm k',\bm q)=  \nn\\
& \langle \Phi_0(Q)| \gamma_{-\bm k + \half \bm q,\alpha} \gamma_{\bm k + \half \bm q,\beta}
    \gamma_{-\bm k'-\half \bm q,\alpha'} \gamma_{\bm k'-  \half \bm q,\beta'}|\Phi_0(Q)\rangle.
\end{align}
Equations~(\ref{Channel1}), (\ref{Channel2}) and (\ref{Channel3}) apply to this four-point function as well, but with the modified definition of $U(\b{k})$. After lengthy calculations, we obtain the interaction variational energy as
\begin{align*}
\langle \Phi_0(Q)| V| \Phi_0(Q)\rangle&=-\frac{g}{32} \sum_{\bm p_1 \bm p_2}
\frac{v}{E_Q(\bm p_1)} \frac{v}{E_Q(\bm p_2)}\\
 &\qquad\times \left[(k_z^2-k_y^2) (\alpha^2 p_{1y} p_{2y} - \beta^2 p_{1z} p_{2z}) \right.\nn\\
&\qquad\quad  \left.-2 k_z k_y \alpha\beta (p_{1y} p_{2z} + p_{1z} p_{2y}) \right],
\end{align*}
where $\bm k \equiv \bm p_1 - \bm p_2$. Performing a change of variables,
\begin{align}
 & p_{1y} = |\b{p}_1| \cos\varphi_1 ,\quad p_{1z} = |\b{p}_1| \sin\varphi_1 \nn\\
 & p_{2y} = |\b{p}_2| \cos\varphi_2 ,\quad p_{2z} = |\b{p}_2| \sin\varphi_2,\nn
\end{align}
we obtain
\begin{align}
&\langle \Phi_0(Q)| V| \Phi_0(Q)\rangle
=-\frac{g}{16} \frac{1}{1+\delta^2} \sum_{\bm p_1 \bm p_2}
 |\b{p}_1||\b{p}_2| \nn\\
&\qquad\times \frac{1+ \delta (\cos2\varphi_1+ \cos 2\varphi_2)
+ \delta^2\cos2\varphi_1 \cos2\varphi_2 }{(1+\delta_1\cos2\varphi_1)^{1/2}(1+\delta_1\cos2\varphi_2)^{1/2}},
\end{align}
where we define $\delta_1\equiv 2\delta/(1+\delta^2)$. Once again the angular integrals can be performed with the use of the elliptic integrals (\ref{EllipticK}) and (\ref{EllipticE}), and we obtain
\begin{align}\label{EVQ}
&\langle \Phi_0(Q)| V| \Phi_0(Q)\rangle
  = - \frac{g}{16\pi^2}\left(\sum_{\bm p} |\b{p}| \right)^2\nn\\
  &\qquad\qquad\times \left[(1+\delta) E(\delta_2)+(1-\delta) K(\delta_2)\right]^2.
\end{align}
Adding the contributions (\ref{E0Q}) and (\ref{EVQ}) and performing the remaining momentum integrals with a large momentum cutoff $\Lambda$, we obtain the total variational energy as
\begin{align}
E_\textrm{MF}(Q)=\frac{vL_\parallel^2\Lambda^3}{6\pi^2}\c{E}_Q(\delta,\alpha),\label{EMFQ}
\end{align}
where the dimensionless function $\c{E}_Q(\delta,\alpha)$ is defined as
\begin{align}
\c{E}_Q(\delta,\alpha)&=\frac{\delta^2}{1+\delta}K(\delta_2)-(1+\delta)E(\delta_2)\nn\\
&\quad-\alpha\left[(1-\delta)K(\delta_2)+(1+\delta)E(\delta_2)\right]^2,\label{dimless_eq}
\end{align}
and $\alpha\equiv gL_\parallel^2\Lambda^3/96\pi^2 v$ is a dimensionless coupling constant. Plotting $\c{E}_Q(\delta,\alpha)$ as a function of the dimensionless nematic order parameter $\delta$ for several values of $\alpha$, we find that there is a continuous transition at a certain critical value of $\alpha=\alpha_c$ (Fig.~\ref{fig:nematic}). To find $\alpha_c$, we expand $\Delta\c{E}_Q(\delta,\alpha)\equiv\c{E}_Q(\delta,\alpha)-\c{E}_Q(0,\alpha)$ in powers of $\delta$,
\begin{align}
\Delta\c{E}_Q(\delta,\alpha)=r\delta^2+u\delta^4+\c{O}(\delta^6),
\end{align}
where
\begin{align}
r=\frac{\pi}{4}\left(\frac{3}{2}+\alpha\right),\quad
u=\frac{3\pi}{64}\left(\frac{5}{2}+\alpha\right).
\end{align}
We see that $r$ changes sign at $\alpha_c=-\frac{3}{2}$, while $u$ remains positive. Hence there is a continuous transition at the negative critical coupling constant
\begin{align}
g_c=-\frac{144\pi^2v}{L_\parallel^2\Lambda^3},
\end{align}
such that $Q=0$ for $g>g_c$ and $Q\neq 0$ for $g<g_c$. There is no nematic instability for $g>0$. Our calculation reveals that bulk Goldstone modes induce a positive coupling constant [Eq.~(\ref{g0simplified})], thus we conclude that the possibility of a surface nematic instability mediated by bulk Goldstone modes in $^3$He-$B$ is unlikely.


\begin{thebibliography}{56}%
\makeatletter
\providecommand \@ifxundefined [1]{%
 \@ifx{#1\undefined}
}%
\providecommand \@ifnum [1]{%
 \ifnum #1\expandafter \@firstoftwo
 \else \expandafter \@secondoftwo
 \fi
}%
\providecommand \@ifx [1]{%
 \ifx #1\expandafter \@firstoftwo
 \else \expandafter \@secondoftwo
 \fi
}%
\providecommand \natexlab [1]{#1}%
\providecommand \enquote  [1]{``#1''}%
\providecommand \bibnamefont  [1]{#1}%
\providecommand \bibfnamefont [1]{#1}%
\providecommand \citenamefont [1]{#1}%
\providecommand \href@noop [0]{\@secondoftwo}%
\providecommand \href [0]{\begingroup \@sanitize@url \@href}%
\providecommand \@href[1]{\@@startlink{#1}\@@href}%
\providecommand \@@href[1]{\endgroup#1\@@endlink}%
\providecommand \@sanitize@url [0]{\catcode `\\12\catcode `\$12\catcode
  `\&12\catcode `\#12\catcode `\^12\catcode `\_12\catcode `\%12\relax}%
\providecommand \@@startlink[1]{}%
\providecommand \@@endlink[0]{}%
\providecommand \url  [0]{\begingroup\@sanitize@url \@url }%
\providecommand \@url [1]{\endgroup\@href {#1}{\urlprefix }}%
\providecommand \urlprefix  [0]{URL }%
\providecommand \Eprint [0]{\href }%
\providecommand \doibase [0]{http://dx.doi.org/}%
\providecommand \selectlanguage [0]{\@gobble}%
\providecommand \bibinfo  [0]{\@secondoftwo}%
\providecommand \bibfield  [0]{\@secondoftwo}%
\providecommand \translation [1]{[#1]}%
\providecommand \BibitemOpen [0]{}%
\providecommand \bibitemStop [0]{}%
\providecommand \bibitemNoStop [0]{.\EOS\space}%
\providecommand \EOS [0]{\spacefactor3000\relax}%
\providecommand \BibitemShut  [1]{\csname bibitem#1\endcsname}%
\let\auto@bib@innerbib\@empty
\bibitem [{\citenamefont {Hasan}\ and\ \citenamefont {Kane}(2010)}]{hasan2010}%
  \BibitemOpen
  \bibfield  {author} {\bibinfo {author} {\bibfnamefont {M.~Z.}\ \bibnamefont
  {Hasan}}\ and\ \bibinfo {author} {\bibfnamefont {C.~L.}\ \bibnamefont
  {Kane}},\ }\href {\doibase 10.1103/RevModPhys.82.3045} {\bibfield  {journal}
  {\bibinfo  {journal} {Rev. Mod. Phys.}\ }\textbf {\bibinfo {volume} {82}},\
  \bibinfo {pages} {3045} (\bibinfo {year} {2010})}\BibitemShut {NoStop}%
\bibitem [{\citenamefont {Qi}\ and\ \citenamefont {Zhang}(2011)}]{qi2011}%
  \BibitemOpen
  \bibfield  {author} {\bibinfo {author} {\bibfnamefont {X.-L.}\ \bibnamefont
  {Qi}}\ and\ \bibinfo {author} {\bibfnamefont {S.-C.}\ \bibnamefont {Zhang}},\
  }\href {\doibase 10.1103/RevModPhys.83.1057} {\bibfield  {journal} {\bibinfo
  {journal} {Rev. Mod. Phys.}\ }\textbf {\bibinfo {volume} {83}},\ \bibinfo
  {pages} {1057} (\bibinfo {year} {2011})}\BibitemShut {NoStop}%
\bibitem [{\citenamefont {Roy}(2006)}]{roy2006}%
  \BibitemOpen
  \bibfield  {author} {\bibinfo {author} {\bibfnamefont {R.}~\bibnamefont
  {Roy}},\ }\href@noop {} {\  (\bibinfo {year} {2006})},\ \Eprint
  {http://arxiv.org/abs/cond-mat/0608064} {arXiv:cond-mat/0608064} \BibitemShut
  {NoStop}%
\bibitem [{\citenamefont {Roy}()}]{roy2008}%
  \BibitemOpen
  \bibfield  {author} {\bibinfo {author} {\bibfnamefont {R.}~\bibnamefont
  {Roy}},\ }\href@noop {} {\ }\Eprint {http://arxiv.org/abs/0803.2868}
  {arXiv:0803.2868} \BibitemShut {NoStop}%
\bibitem [{\citenamefont {Schnyder}\ \emph {et~al.}(2008)\citenamefont
  {Schnyder}, \citenamefont {Ryu}, \citenamefont {Furusaki},\ and\
  \citenamefont {Ludwig}}]{schnyder2008}%
  \BibitemOpen
  \bibfield  {author} {\bibinfo {author} {\bibfnamefont {A.~P.}\ \bibnamefont
  {Schnyder}}, \bibinfo {author} {\bibfnamefont {S.}~\bibnamefont {Ryu}},
  \bibinfo {author} {\bibfnamefont {A.}~\bibnamefont {Furusaki}}, \ and\
  \bibinfo {author} {\bibfnamefont {A.~W.~W.}\ \bibnamefont {Ludwig}},\ }\href
  {\doibase 10.1103/PhysRevB.78.195125} {\bibfield  {journal} {\bibinfo
  {journal} {Phys. Rev. B}\ }\textbf {\bibinfo {volume} {78}},\ \bibinfo
  {pages} {195125} (\bibinfo {year} {2008})}\BibitemShut {NoStop}%
\bibitem [{\citenamefont {Qi}\ \emph {et~al.}(2009)\citenamefont {Qi},
  \citenamefont {Hughes}, \citenamefont {Raghu},\ and\ \citenamefont
  {Zhang}}]{qi2009}%
  \BibitemOpen
  \bibfield  {author} {\bibinfo {author} {\bibfnamefont {X.-L.}\ \bibnamefont
  {Qi}}, \bibinfo {author} {\bibfnamefont {T.~L.}\ \bibnamefont {Hughes}},
  \bibinfo {author} {\bibfnamefont {S.}~\bibnamefont {Raghu}}, \ and\ \bibinfo
  {author} {\bibfnamefont {S.-C.}\ \bibnamefont {Zhang}},\ }\href {\doibase
  10.1103/PhysRevLett.102.187001} {\bibfield  {journal} {\bibinfo  {journal}
  {Phys. Rev. Lett.}\ }\textbf {\bibinfo {volume} {102}},\ \bibinfo {pages}
  {187001} (\bibinfo {year} {2009})}\BibitemShut {NoStop}%
\bibitem [{\citenamefont {Schnyder}\ \emph {et~al.}(2009)\citenamefont
  {Schnyder}, \citenamefont {Ryu}, \citenamefont {Furusaki},\ and\
  \citenamefont {Ludwig}}]{schnyder2009}%
  \BibitemOpen
  \bibfield  {author} {\bibinfo {author} {\bibfnamefont {A.~P.}\ \bibnamefont
  {Schnyder}}, \bibinfo {author} {\bibfnamefont {S.}~\bibnamefont {Ryu}},
  \bibinfo {author} {\bibfnamefont {A.}~\bibnamefont {Furusaki}}, \ and\
  \bibinfo {author} {\bibfnamefont {A.~W.~W.}\ \bibnamefont {Ludwig}},\ }\href
  {\doibase 10.1063/1.3149481} {\bibfield  {journal} {\bibinfo  {journal} {AIP
  Conf. Proc.}\ }\textbf {\bibinfo {volume} {1134}},\ \bibinfo {pages} {10}
  (\bibinfo {year} {2009})}\BibitemShut {NoStop}%
\bibitem [{\citenamefont {Kitaev}(2009)}]{kitaev2009}%
  \BibitemOpen
  \bibfield  {author} {\bibinfo {author} {\bibfnamefont {A.}~\bibnamefont
  {Kitaev}},\ }\href {\doibase 10.1063/1.3149495} {\bibfield  {journal}
  {\bibinfo  {journal} {AIP Conf. Proc.}\ }\textbf {\bibinfo {volume} {1134}},\
  \bibinfo {pages} {22} (\bibinfo {year} {2009})}\BibitemShut {NoStop}%
\bibitem [{\citenamefont {Wilczek}(2009)}]{wilczek2009}%
  \BibitemOpen
  \bibfield  {author} {\bibinfo {author} {\bibfnamefont {F.}~\bibnamefont
  {Wilczek}},\ }\href {\doibase 10.1038/nphys1380} {\bibfield  {journal}
  {\bibinfo  {journal} {Nature Phys.}\ }\textbf {\bibinfo {volume} {5}},\
  \bibinfo {pages} {614} (\bibinfo {year} {2009})}\BibitemShut {NoStop}%
\bibitem [{\citenamefont {Nayak}\ \emph {et~al.}(2008)\citenamefont {Nayak},
  \citenamefont {Simon}, \citenamefont {Stern}, \citenamefont {Freedman},\ and\
  \citenamefont {Das~Sarma}}]{nayak2008}%
  \BibitemOpen
  \bibfield  {author} {\bibinfo {author} {\bibfnamefont {C.}~\bibnamefont
  {Nayak}}, \bibinfo {author} {\bibfnamefont {S.~H.}\ \bibnamefont {Simon}},
  \bibinfo {author} {\bibfnamefont {A.}~\bibnamefont {Stern}}, \bibinfo
  {author} {\bibfnamefont {M.}~\bibnamefont {Freedman}}, \ and\ \bibinfo
  {author} {\bibfnamefont {S.}~\bibnamefont {Das~Sarma}},\ }\href {\doibase
  10.1103/RevModPhys.80.1083} {\bibfield  {journal} {\bibinfo  {journal} {Rev.
  Mod. Phys.}\ }\textbf {\bibinfo {volume} {80}},\ \bibinfo {pages} {1083}
  (\bibinfo {year} {2008})}\BibitemShut {NoStop}%
\bibitem [{\citenamefont {Alicea}(2012)}]{alicea2012}%
  \BibitemOpen
  \bibfield  {author} {\bibinfo {author} {\bibfnamefont {J.}~\bibnamefont
  {Alicea}},\ }\href {\doibase 10.1088/0034-4885/75/7/076501} {\bibfield
  {journal} {\bibinfo  {journal} {Rep. Prog. Phys.}\ }\textbf {\bibinfo
  {volume} {75}},\ \bibinfo {pages} {076501} (\bibinfo {year}
  {2012})}\BibitemShut {NoStop}%
\bibitem [{\citenamefont {Volovik}(2003)}]{VolovikBook}%
  \BibitemOpen
  \bibfield  {author} {\bibinfo {author} {\bibfnamefont {G.~E.}\ \bibnamefont
  {Volovik}},\ }\href@noop {} {\emph {\bibinfo {title} {The Universe in a
  Helium Droplet}}}\ (\bibinfo  {publisher} {Oxford University Press},\
  \bibinfo {address} {Oxford},\ \bibinfo {year} {2003})\BibitemShut {NoStop}%
\bibitem [{\citenamefont {Salomaa}\ and\ \citenamefont
  {Volovik}(1988)}]{salomaa1988}%
  \BibitemOpen
  \bibfield  {author} {\bibinfo {author} {\bibfnamefont {M.~M.}\ \bibnamefont
  {Salomaa}}\ and\ \bibinfo {author} {\bibfnamefont {G.~E.}\ \bibnamefont
  {Volovik}},\ }\href {\doibase 10.1103/PhysRevB.37.9298} {\bibfield  {journal}
  {\bibinfo  {journal} {Phys. Rev. B}\ }\textbf {\bibinfo {volume} {37}},\
  \bibinfo {pages} {9298} (\bibinfo {year} {1988})}\BibitemShut {NoStop}%
\bibitem [{\citenamefont {Balian}\ and\ \citenamefont
  {Werthamer}(1963)}]{balian1963}%
  \BibitemOpen
  \bibfield  {author} {\bibinfo {author} {\bibfnamefont {R.}~\bibnamefont
  {Balian}}\ and\ \bibinfo {author} {\bibfnamefont {N.~R.}\ \bibnamefont
  {Werthamer}},\ }\href {\doibase 10.1103/PhysRev.131.1553} {\bibfield
  {journal} {\bibinfo  {journal} {Phys. Rev.}\ }\textbf {\bibinfo {volume}
  {131}},\ \bibinfo {pages} {1553} (\bibinfo {year} {1963})}\BibitemShut
  {NoStop}%
\bibitem [{vdo()}]{vdovin1963}%
  \BibitemOpen
  \href@noop {} {}\bibinfo {note} {Yu. A. Vdovin, in {\it Applications of the
  Methods of Quantum Field Theory to the Many-Body Problem}, edited by A. I.
  Alekseyeva (Gosatomizdat, Moscow, 1963).}\BibitemShut {Stop}%
\bibitem [{\citenamefont {Volovik}(2009)}]{volovik2009}%
  \BibitemOpen
  \bibfield  {author} {\bibinfo {author} {\bibfnamefont {G.~E.}\ \bibnamefont
  {Volovik}},\ }\href {\doibase 10.1134/S0021364009170172} {\bibfield
  {journal} {\bibinfo  {journal} {{JETP} Lett.}\ }\textbf {\bibinfo {volume}
  {90}},\ \bibinfo {pages} {398} (\bibinfo {year} {2009})}\BibitemShut
  {NoStop}%
\bibitem [{\citenamefont {Grover}\ and\ \citenamefont
  {Vishwanath}()}]{grover2012}%
  \BibitemOpen
  \bibfield  {author} {\bibinfo {author} {\bibfnamefont {T.}~\bibnamefont
  {Grover}}\ and\ \bibinfo {author} {\bibfnamefont {A.}~\bibnamefont
  {Vishwanath}},\ }\href@noop {} {\ }\Eprint {http://arxiv.org/abs/1206.1332}
  {arXiv:1206.1332} \BibitemShut {NoStop}%
\bibitem [{\citenamefont {Li}\ \emph {et~al.}(2009)\citenamefont {Li},
  \citenamefont {Belitz},\ and\ \citenamefont {Toner}}]{li2009}%
  \BibitemOpen
  \bibfield  {author} {\bibinfo {author} {\bibfnamefont {Q.}~\bibnamefont
  {Li}}, \bibinfo {author} {\bibfnamefont {D.}~\bibnamefont {Belitz}}, \ and\
  \bibinfo {author} {\bibfnamefont {J.}~\bibnamefont {Toner}},\ }\href
  {\doibase 10.1103/PhysRevB.79.054514} {\bibfield  {journal} {\bibinfo
  {journal} {Phys. Rev. B}\ }\textbf {\bibinfo {volume} {79}},\ \bibinfo
  {pages} {054514} (\bibinfo {year} {2009})}\BibitemShut {NoStop}%
\bibitem [{\citenamefont {Bauer}\ \emph {et~al.}(2013)\citenamefont {Bauer},
  \citenamefont {Lutchyn}, \citenamefont {Hastings},\ and\ \citenamefont
  {Troyer}}]{bauer2013}%
  \BibitemOpen
  \bibfield  {author} {\bibinfo {author} {\bibfnamefont {B.}~\bibnamefont
  {Bauer}}, \bibinfo {author} {\bibfnamefont {R.~M.}\ \bibnamefont {Lutchyn}},
  \bibinfo {author} {\bibfnamefont {M.~B.}\ \bibnamefont {Hastings}}, \ and\
  \bibinfo {author} {\bibfnamefont {M.}~\bibnamefont {Troyer}},\ }\href
  {\doibase 10.1103/PhysRevB.87.014503} {\bibfield  {journal} {\bibinfo
  {journal} {Phys. Rev. B}\ }\textbf {\bibinfo {volume} {87}},\ \bibinfo
  {pages} {014503} (\bibinfo {year} {2013})}\BibitemShut {NoStop}%
\bibitem [{\citenamefont {Fidkowski}\ \emph {et~al.}(2011)\citenamefont
  {Fidkowski}, \citenamefont {Lutchyn}, \citenamefont {Nayak},\ and\
  \citenamefont {Fisher}}]{fidkowski2011}%
  \BibitemOpen
  \bibfield  {author} {\bibinfo {author} {\bibfnamefont {L.}~\bibnamefont
  {Fidkowski}}, \bibinfo {author} {\bibfnamefont {R.~M.}\ \bibnamefont
  {Lutchyn}}, \bibinfo {author} {\bibfnamefont {C.}~\bibnamefont {Nayak}}, \
  and\ \bibinfo {author} {\bibfnamefont {M.~P.~A.}\ \bibnamefont {Fisher}},\
  }\href {\doibase 10.1103/PhysRevB.84.195436} {\bibfield  {journal} {\bibinfo
  {journal} {Phys. Rev. B}\ }\textbf {\bibinfo {volume} {84}},\ \bibinfo
  {pages} {195436} (\bibinfo {year} {2011})}\BibitemShut {NoStop}%
\bibitem [{Sco()}]{ScottThomas}%
  \BibitemOpen
  \href@noop {} {}\bibinfo {note} {S. Thomas, talk at the 2005 KITP Conference
  on Quantum Phase Transitions, Kavli Institute for Theoretical Physics, Santa
  Barbara, 21 January 2005.}\BibitemShut {Stop}%
\bibitem [{\citenamefont {Sonoda}(2011)}]{sonoda2011}%
  \BibitemOpen
  \bibfield  {author} {\bibinfo {author} {\bibfnamefont {H.}~\bibnamefont
  {Sonoda}},\ }\href {\doibase 10.1143/PTP.126.57} {\bibfield  {journal}
  {\bibinfo  {journal} {Prog. Theor. Phys.}\ }\textbf {\bibinfo {volume}
  {126}},\ \bibinfo {pages} {57} (\bibinfo {year} {2011})}\BibitemShut
  {NoStop}%
\bibitem [{\citenamefont {Grover}\ \emph {et~al.}(2014)\citenamefont {Grover},
  \citenamefont {Sheng},\ and\ \citenamefont {Vishwanath}}]{grover2014}%
  \BibitemOpen
  \bibfield  {author} {\bibinfo {author} {\bibfnamefont {T.}~\bibnamefont
  {Grover}}, \bibinfo {author} {\bibfnamefont {D.~N.}\ \bibnamefont {Sheng}}, \
  and\ \bibinfo {author} {\bibfnamefont {A.}~\bibnamefont {Vishwanath}},\
  }\href {\doibase 10.1126/science.1248253} {\bibfield  {journal} {\bibinfo
  {journal} {Science}\ }\textbf {\bibinfo {volume} {344}},\ \bibinfo {pages}
  {280} (\bibinfo {year} {2014})}\BibitemShut {NoStop}%
\bibitem [{\citenamefont {Okuda}\ and\ \citenamefont
  {Nomura}(2012)}]{okuda2012}%
  \BibitemOpen
  \bibfield  {author} {\bibinfo {author} {\bibfnamefont {Y.}~\bibnamefont
  {Okuda}}\ and\ \bibinfo {author} {\bibfnamefont {R.}~\bibnamefont {Nomura}},\
  }\href {\doibase 10.1088/0953-8984/24/34/343201} {\bibfield  {journal}
  {\bibinfo  {journal} {J. Phys.: Condens. Matter}\ }\textbf {\bibinfo {volume}
  {24}},\ \bibinfo {pages} {343201} (\bibinfo {year} {2012})}\BibitemShut
  {NoStop}%
\bibitem [{\citenamefont {Vollhardt}\ and\ \citenamefont {{P.
  W\"{o}lfle}}(2013)}]{VollhardtWolfle}%
  \BibitemOpen
  \bibfield  {author} {\bibinfo {author} {\bibfnamefont {D.}~\bibnamefont
  {Vollhardt}}\ and\ \bibinfo {author} {\bibnamefont {{P. W\"{o}lfle}}},\
  }\href@noop {} {\emph {\bibinfo {title} {The Superfluid Phases of Helium
  3}}}\ (\bibinfo  {publisher} {Dover},\ \bibinfo {address} {New York},\
  \bibinfo {year} {2013})\BibitemShut {NoStop}%
\bibitem [{\citenamefont {Leggett}(1973)}]{leggett1973}%
  \BibitemOpen
  \bibfield  {author} {\bibinfo {author} {\bibfnamefont {A.~J.}\ \bibnamefont
  {Leggett}},\ }\href {\doibase 10.1103/PhysRevLett.31.352} {\bibfield
  {journal} {\bibinfo  {journal} {Phys. Rev. Lett.}\ }\textbf {\bibinfo
  {volume} {31}},\ \bibinfo {pages} {352} (\bibinfo {year} {1973})}\BibitemShut
  {NoStop}%
\bibitem [{\citenamefont {Leggett}(1974)}]{leggett1974}%
  \BibitemOpen
  \bibfield  {author} {\bibinfo {author} {\bibfnamefont {A.~J.}\ \bibnamefont
  {Leggett}},\ }\href {\doibase 10.1016/0003-4916(74)90277-2} {\bibfield
  {journal} {\bibinfo  {journal} {Ann. Phys. (N.Y.)}\ }\textbf {\bibinfo
  {volume} {85}},\ \bibinfo {pages} {11} (\bibinfo {year} {1974})}\BibitemShut
  {NoStop}%
\bibitem [{\citenamefont {Brinkman}\ and\ \citenamefont
  {Smith}(1974)}]{brinkman1974b}%
  \BibitemOpen
  \bibfield  {author} {\bibinfo {author} {\bibfnamefont {W.~F.}\ \bibnamefont
  {Brinkman}}\ and\ \bibinfo {author} {\bibfnamefont {H.}~\bibnamefont
  {Smith}},\ }\href {\doibase 10.1103/PhysRevA.10.2325} {\bibfield  {journal}
  {\bibinfo  {journal} {Phys. Rev. A}\ }\textbf {\bibinfo {volume} {10}},\
  \bibinfo {pages} {2325} (\bibinfo {year} {1974})}\BibitemShut {NoStop}%
\bibitem [{\citenamefont {Smith}\ \emph {et~al.}(1977)\citenamefont {Smith},
  \citenamefont {Brinkman},\ and\ \citenamefont {Engelsberg}}]{smith1977}%
  \BibitemOpen
  \bibfield  {author} {\bibinfo {author} {\bibfnamefont {H.}~\bibnamefont
  {Smith}}, \bibinfo {author} {\bibfnamefont {W.~F.}\ \bibnamefont {Brinkman}},
  \ and\ \bibinfo {author} {\bibfnamefont {S.}~\bibnamefont {Engelsberg}},\
  }\href {\doibase 10.1103/PhysRevB.15.199} {\bibfield  {journal} {\bibinfo
  {journal} {Phys. Rev. B}\ }\textbf {\bibinfo {volume} {15}},\ \bibinfo
  {pages} {199} (\bibinfo {year} {1977})}\BibitemShut {NoStop}%
\bibitem [{\citenamefont {Nagato}\ \emph {et~al.}(2009)\citenamefont {Nagato},
  \citenamefont {Higashitani},\ and\ \citenamefont {Nagai}}]{nagato2009}%
  \BibitemOpen
  \bibfield  {author} {\bibinfo {author} {\bibfnamefont {Y.}~\bibnamefont
  {Nagato}}, \bibinfo {author} {\bibfnamefont {S.}~\bibnamefont {Higashitani}},
  \ and\ \bibinfo {author} {\bibfnamefont {K.}~\bibnamefont {Nagai}},\ }\href
  {\doibase 10.1143/JPSJ.78.123603} {\bibfield  {journal} {\bibinfo  {journal}
  {J. Phys. Soc. Jpn.}\ }\textbf {\bibinfo {volume} {78}},\ \bibinfo {pages}
  {123603} (\bibinfo {year} {2009})}\BibitemShut {NoStop}%
\bibitem [{\citenamefont {Chung}\ and\ \citenamefont
  {Zhang}(2009)}]{chung2009}%
  \BibitemOpen
  \bibfield  {author} {\bibinfo {author} {\bibfnamefont {S.~B.}\ \bibnamefont
  {Chung}}\ and\ \bibinfo {author} {\bibfnamefont {S.-C.}\ \bibnamefont
  {Zhang}},\ }\href {\doibase 10.1103/PhysRevLett.103.235301} {\bibfield
  {journal} {\bibinfo  {journal} {Phys. Rev. Lett.}\ }\textbf {\bibinfo
  {volume} {103}},\ \bibinfo {pages} {235301} (\bibinfo {year}
  {2009})}\BibitemShut {NoStop}%
\bibitem [{\citenamefont {Vorontsov}\ and\ \citenamefont
  {Sauls}(2003)}]{vorontsov2003}%
  \BibitemOpen
  \bibfield  {author} {\bibinfo {author} {\bibfnamefont {A.~B.}\ \bibnamefont
  {Vorontsov}}\ and\ \bibinfo {author} {\bibfnamefont {J.~A.}\ \bibnamefont
  {Sauls}},\ }\href {\doibase 10.1103/PhysRevB.68.064508} {\bibfield  {journal}
  {\bibinfo  {journal} {Phys. Rev. B}\ }\textbf {\bibinfo {volume} {68}},\
  \bibinfo {pages} {064508} (\bibinfo {year} {2003})}\BibitemShut {NoStop}%
\bibitem [{\citenamefont {{W\"{o}lfle}}(1977)}]{wolfle1977}%
  \BibitemOpen
  \bibfield  {author} {\bibinfo {author} {\bibfnamefont {P.}~\bibnamefont
  {{W\"{o}lfle}}},\ }\href {\doibase 10.1016/0378-4363(77)90015-8} {\bibfield
  {journal} {\bibinfo  {journal} {Physica B}\ }\textbf {\bibinfo {volume}
  {90}},\ \bibinfo {pages} {96} (\bibinfo {year} {1977})}\BibitemShut {NoStop}%
\bibitem [{\citenamefont {Adler}(1965{\natexlab{a}})}]{adler1965a}%
  \BibitemOpen
  \bibfield  {author} {\bibinfo {author} {\bibfnamefont {S.~L.}\ \bibnamefont
  {Adler}},\ }\href {\doibase 10.1103/PhysRev.137.B1022} {\bibfield  {journal}
  {\bibinfo  {journal} {Phys. Rev.}\ }\textbf {\bibinfo {volume} {137}},\
  \bibinfo {pages} {B1022} (\bibinfo {year} {1965}{\natexlab{a}})}\BibitemShut
  {NoStop}%
\bibitem [{\citenamefont {Adler}(1965{\natexlab{b}})}]{adler1965b}%
  \BibitemOpen
  \bibfield  {author} {\bibinfo {author} {\bibfnamefont {S.~L.}\ \bibnamefont
  {Adler}},\ }\href {\doibase 10.1103/PhysRev.139.B1638} {\bibfield  {journal}
  {\bibinfo  {journal} {Phys. Rev.}\ }\textbf {\bibinfo {volume} {139}},\
  \bibinfo {pages} {B1638} (\bibinfo {year} {1965}{\natexlab{b}})}\BibitemShut
  {NoStop}%
\bibitem [{\citenamefont {Witczak-Krempa}\ \emph {et~al.}(2010)\citenamefont
  {Witczak-Krempa}, \citenamefont {Choy},\ and\ \citenamefont
  {Kim}}]{witczak-krempa2010}%
  \BibitemOpen
  \bibfield  {author} {\bibinfo {author} {\bibfnamefont {W.}~\bibnamefont
  {Witczak-Krempa}}, \bibinfo {author} {\bibfnamefont {T.~P.}\ \bibnamefont
  {Choy}}, \ and\ \bibinfo {author} {\bibfnamefont {Y.~B.}\ \bibnamefont
  {Kim}},\ }\href {\doibase 10.1103/PhysRevB.82.165122} {\bibfield  {journal}
  {\bibinfo  {journal} {Phys. Rev. B}\ }\textbf {\bibinfo {volume} {82}},\
  \bibinfo {pages} {165122} (\bibinfo {year} {2010})}\BibitemShut {NoStop}%
\bibitem [{\citenamefont {Habe}(2014)}]{habe2014}%
  \BibitemOpen
  \bibfield  {author} {\bibinfo {author} {\bibfnamefont {T.}~\bibnamefont
  {Habe}},\ }\href {\doibase 10.1103/PhysRevB.89.035305} {\bibfield  {journal}
  {\bibinfo  {journal} {Phys. Rev. B}\ }\textbf {\bibinfo {volume} {89}},\
  \bibinfo {pages} {035305} (\bibinfo {year} {2014})}\BibitemShut {NoStop}%
\bibitem [{\citenamefont {{de Gennes}}\ and\ \citenamefont
  {Rainer}(1974)}]{DeGennes1974}%
  \BibitemOpen
  \bibfield  {author} {\bibinfo {author} {\bibfnamefont {P.~G.}\ \bibnamefont
  {{de Gennes}}}\ and\ \bibinfo {author} {\bibfnamefont {D.}~\bibnamefont
  {Rainer}},\ }\href {\doibase 10.1016/0375-9601(74)90949-9} {\bibfield
  {journal} {\bibinfo  {journal} {Phys. Lett. A}\ }\textbf {\bibinfo {volume}
  {46}},\ \bibinfo {pages} {429} (\bibinfo {year} {1974})}\BibitemShut
  {NoStop}%
\bibitem [{\citenamefont {W{\"o}lfle}(1974)}]{wolfle1974}%
  \BibitemOpen
  \bibfield  {author} {\bibinfo {author} {\bibfnamefont {P.}~\bibnamefont
  {W{\"o}lfle}},\ }\href {\doibase 10.1016/0375-9601(74)90018-8} {\bibfield
  {journal} {\bibinfo  {journal} {Phys. Lett. A}\ }\textbf {\bibinfo {volume}
  {47}},\ \bibinfo {pages} {224} (\bibinfo {year} {1974})}\BibitemShut
  {NoStop}%
\bibitem [{\citenamefont {Brinkman}\ and\ \citenamefont
  {Cross}(1978)}]{brinkman1978}%
  \BibitemOpen
  \bibfield  {author} {\bibinfo {author} {\bibfnamefont {W.~F.}\ \bibnamefont
  {Brinkman}}\ and\ \bibinfo {author} {\bibfnamefont {M.~C.}\ \bibnamefont
  {Cross}},\ }\href@noop {} {\emph {\bibinfo {title} {Progress in
  Low-Temperature Physics: Dynamics of Superfluid Helium 3}}}\ (\bibinfo
  {publisher} {North-Holland},\ \bibinfo {address} {New York},\ \bibinfo {year}
  {1978})\BibitemShut {NoStop}%
\bibitem [{\citenamefont {Murakawa}\ \emph
  {et~al.}(2009{\natexlab{a}})\citenamefont {Murakawa}, \citenamefont {Tamura},
  \citenamefont {Wada}, \citenamefont {Wasai}, \citenamefont {Saitoh},
  \citenamefont {Aoki}, \citenamefont {Nomura}, \citenamefont {Okuda},
  \citenamefont {Nagato}, \citenamefont {Yamamoto}, \citenamefont
  {Higashitani},\ and\ \citenamefont {Nagai}}]{murakawa2009}%
  \BibitemOpen
  \bibfield  {author} {\bibinfo {author} {\bibfnamefont {S.}~\bibnamefont
  {Murakawa}}, \bibinfo {author} {\bibfnamefont {Y.}~\bibnamefont {Tamura}},
  \bibinfo {author} {\bibfnamefont {Y.}~\bibnamefont {Wada}}, \bibinfo {author}
  {\bibfnamefont {M.}~\bibnamefont {Wasai}}, \bibinfo {author} {\bibfnamefont
  {M.}~\bibnamefont {Saitoh}}, \bibinfo {author} {\bibfnamefont
  {Y.}~\bibnamefont {Aoki}}, \bibinfo {author} {\bibfnamefont {R.}~\bibnamefont
  {Nomura}}, \bibinfo {author} {\bibfnamefont {Y.}~\bibnamefont {Okuda}},
  \bibinfo {author} {\bibfnamefont {Y.}~\bibnamefont {Nagato}}, \bibinfo
  {author} {\bibfnamefont {M.}~\bibnamefont {Yamamoto}}, \bibinfo {author}
  {\bibfnamefont {S.}~\bibnamefont {Higashitani}}, \ and\ \bibinfo {author}
  {\bibfnamefont {K.}~\bibnamefont {Nagai}},\ }\href {\doibase
  10.1103/PhysRevLett.103.155301} {\bibfield  {journal} {\bibinfo  {journal}
  {Phys. Rev. Lett.}\ }\textbf {\bibinfo {volume} {103}},\ \bibinfo {pages}
  {155301} (\bibinfo {year} {2009}{\natexlab{a}})}\BibitemShut {NoStop}%
\bibitem [{\citenamefont {Tsutsumi}\ \emph {et~al.}(2011)\citenamefont
  {Tsutsumi}, \citenamefont {Ichioka},\ and\ \citenamefont
  {Machida}}]{tsutsumi2011}%
  \BibitemOpen
  \bibfield  {author} {\bibinfo {author} {\bibfnamefont {Y.}~\bibnamefont
  {Tsutsumi}}, \bibinfo {author} {\bibfnamefont {M.}~\bibnamefont {Ichioka}}, \
  and\ \bibinfo {author} {\bibfnamefont {K.}~\bibnamefont {Machida}},\ }\href
  {\doibase 10.1103/PhysRevB.83.094510} {\bibfield  {journal} {\bibinfo
  {journal} {Phys. Rev. B}\ }\textbf {\bibinfo {volume} {83}},\ \bibinfo
  {pages} {094510} (\bibinfo {year} {2011})}\BibitemShut {NoStop}%
\bibitem [{\citenamefont {Fidkowski}\ \emph {et~al.}(2013)\citenamefont
  {Fidkowski}, \citenamefont {Chen},\ and\ \citenamefont
  {Vishwanath}}]{fidkowski2013}%
  \BibitemOpen
  \bibfield  {author} {\bibinfo {author} {\bibfnamefont {L.}~\bibnamefont
  {Fidkowski}}, \bibinfo {author} {\bibfnamefont {X.}~\bibnamefont {Chen}}, \
  and\ \bibinfo {author} {\bibfnamefont {A.}~\bibnamefont {Vishwanath}},\
  }\href {\doibase 10.1103/PhysRevX.3.041016} {\bibfield  {journal} {\bibinfo
  {journal} {Phys. Rev. X}\ }\textbf {\bibinfo {volume} {3}},\ \bibinfo {pages}
  {041016} (\bibinfo {year} {2013})}\BibitemShut {NoStop}%
\bibitem [{\citenamefont {Metlitski}\ \emph {et~al.}()\citenamefont
  {Metlitski}, \citenamefont {Fidkowski}, \citenamefont {Chen},\ and\
  \citenamefont {Vishwanath}}]{metlitski2014}%
  \BibitemOpen
  \bibfield  {author} {\bibinfo {author} {\bibfnamefont {M.~A.}\ \bibnamefont
  {Metlitski}}, \bibinfo {author} {\bibfnamefont {L.}~\bibnamefont
  {Fidkowski}}, \bibinfo {author} {\bibfnamefont {X.}~\bibnamefont {Chen}}, \
  and\ \bibinfo {author} {\bibfnamefont {A.}~\bibnamefont {Vishwanath}},\
  }\href@noop {} {\ }\Eprint {http://arxiv.org/abs/1406.3032} {arXiv:1406.3032}
  \BibitemShut {NoStop}%
\bibitem [{\citenamefont {{de Gennes}}\ and\ \citenamefont
  {Prost}(1993)}]{DeGennes}%
  \BibitemOpen
  \bibfield  {author} {\bibinfo {author} {\bibfnamefont {P.~G.}\ \bibnamefont
  {{de Gennes}}}\ and\ \bibinfo {author} {\bibfnamefont {J.}~\bibnamefont
  {Prost}},\ }\href@noop {} {\emph {\bibinfo {title} {The Physics of Liquid
  Crystals}}}\ (\bibinfo  {publisher} {Clarendon Press},\ \bibinfo {address}
  {Oxford},\ \bibinfo {year} {1993})\BibitemShut {NoStop}%
\bibitem [{\citenamefont {Wu}\ and\ \citenamefont {Sauls}(2013)}]{wu2013}%
  \BibitemOpen
  \bibfield  {author} {\bibinfo {author} {\bibfnamefont {H.}~\bibnamefont
  {Wu}}\ and\ \bibinfo {author} {\bibfnamefont {J.~A.}\ \bibnamefont {Sauls}},\
  }\href {\doibase 10.1103/PhysRevB.88.184506} {\bibfield  {journal} {\bibinfo
  {journal} {Phys. Rev. B}\ }\textbf {\bibinfo {volume} {88}},\ \bibinfo
  {pages} {184506} (\bibinfo {year} {2013})}\BibitemShut {NoStop}%
\bibitem [{\citenamefont {Aoki}\ \emph {et~al.}(2005)\citenamefont {Aoki},
  \citenamefont {Wada}, \citenamefont {Saitoh}, \citenamefont {Nomura},
  \citenamefont {Okuda}, \citenamefont {Nagato}, \citenamefont {Yamamoto},
  \citenamefont {Higashitani},\ and\ \citenamefont {Nagai}}]{aoki2005}%
  \BibitemOpen
  \bibfield  {author} {\bibinfo {author} {\bibfnamefont {Y.}~\bibnamefont
  {Aoki}}, \bibinfo {author} {\bibfnamefont {Y.}~\bibnamefont {Wada}}, \bibinfo
  {author} {\bibfnamefont {M.}~\bibnamefont {Saitoh}}, \bibinfo {author}
  {\bibfnamefont {R.}~\bibnamefont {Nomura}}, \bibinfo {author} {\bibfnamefont
  {Y.}~\bibnamefont {Okuda}}, \bibinfo {author} {\bibfnamefont
  {Y.}~\bibnamefont {Nagato}}, \bibinfo {author} {\bibfnamefont
  {M.}~\bibnamefont {Yamamoto}}, \bibinfo {author} {\bibfnamefont
  {S.}~\bibnamefont {Higashitani}}, \ and\ \bibinfo {author} {\bibfnamefont
  {K.}~\bibnamefont {Nagai}},\ }\href {\doibase 10.1103/PhysRevLett.95.075301}
  {\bibfield  {journal} {\bibinfo  {journal} {Phys. Rev. Lett.}\ }\textbf
  {\bibinfo {volume} {95}},\ \bibinfo {pages} {075301} (\bibinfo {year}
  {2005})}\BibitemShut {NoStop}%
\bibitem [{\citenamefont {Saitoh}\ \emph {et~al.}(2006)\citenamefont {Saitoh},
  \citenamefont {Wada}, \citenamefont {Aoki}, \citenamefont {Murakawa},
  \citenamefont {Nomura},\ and\ \citenamefont {Okuda}}]{saitoh2006}%
  \BibitemOpen
  \bibfield  {author} {\bibinfo {author} {\bibfnamefont {M.}~\bibnamefont
  {Saitoh}}, \bibinfo {author} {\bibfnamefont {Y.}~\bibnamefont {Wada}},
  \bibinfo {author} {\bibfnamefont {Y.}~\bibnamefont {Aoki}}, \bibinfo {author}
  {\bibfnamefont {S.}~\bibnamefont {Murakawa}}, \bibinfo {author}
  {\bibfnamefont {R.}~\bibnamefont {Nomura}}, \ and\ \bibinfo {author}
  {\bibfnamefont {Y.}~\bibnamefont {Okuda}},\ }\href {\doibase
  10.1103/PhysRevB.74.220505} {\bibfield  {journal} {\bibinfo  {journal} {Phys.
  Rev. B}\ }\textbf {\bibinfo {volume} {74}},\ \bibinfo {pages} {220505}
  (\bibinfo {year} {2006})}\BibitemShut {NoStop}%
\bibitem [{\citenamefont {Nagai}\ \emph {et~al.}(2008)\citenamefont {Nagai},
  \citenamefont {Nagato}, \citenamefont {Yamamoto},\ and\ \citenamefont
  {Higashitani}}]{nagai2008}%
  \BibitemOpen
  \bibfield  {author} {\bibinfo {author} {\bibfnamefont {K.}~\bibnamefont
  {Nagai}}, \bibinfo {author} {\bibfnamefont {Y.}~\bibnamefont {Nagato}},
  \bibinfo {author} {\bibfnamefont {M.}~\bibnamefont {Yamamoto}}, \ and\
  \bibinfo {author} {\bibfnamefont {S.}~\bibnamefont {Higashitani}},\ }\href
  {\doibase 10.1143/JPSJ.77.111003} {\bibfield  {journal} {\bibinfo  {journal}
  {J. Phys. Soc. Jpn.}\ }\textbf {\bibinfo {volume} {77}},\ \bibinfo {pages}
  {111003} (\bibinfo {year} {2008})}\BibitemShut {NoStop}%
\bibitem [{\citenamefont {Wada}\ \emph {et~al.}(2008)\citenamefont {Wada},
  \citenamefont {Murakawa}, \citenamefont {Tamura}, \citenamefont {Saitoh},
  \citenamefont {Aoki}, \citenamefont {Nomura},\ and\ \citenamefont
  {Okuda}}]{wada2008}%
  \BibitemOpen
  \bibfield  {author} {\bibinfo {author} {\bibfnamefont {Y.}~\bibnamefont
  {Wada}}, \bibinfo {author} {\bibfnamefont {S.}~\bibnamefont {Murakawa}},
  \bibinfo {author} {\bibfnamefont {Y.}~\bibnamefont {Tamura}}, \bibinfo
  {author} {\bibfnamefont {M.}~\bibnamefont {Saitoh}}, \bibinfo {author}
  {\bibfnamefont {Y.}~\bibnamefont {Aoki}}, \bibinfo {author} {\bibfnamefont
  {R.}~\bibnamefont {Nomura}}, \ and\ \bibinfo {author} {\bibfnamefont
  {Y.}~\bibnamefont {Okuda}},\ }\href {\doibase 10.1103/PhysRevB.78.214516}
  {\bibfield  {journal} {\bibinfo  {journal} {Phys. Rev. B}\ }\textbf {\bibinfo
  {volume} {78}},\ \bibinfo {pages} {214516} (\bibinfo {year}
  {2008})}\BibitemShut {NoStop}%
\bibitem [{\citenamefont {Murakawa}\ \emph
  {et~al.}(2009{\natexlab{b}})\citenamefont {Murakawa}, \citenamefont {Wada},
  \citenamefont {Tamura}, \citenamefont {Saitoh}, \citenamefont {Aoki},
  \citenamefont {Nomura}, \citenamefont {Okuda}, \citenamefont {Nagato},
  \citenamefont {Yamamoto}, \citenamefont {Higashitani},\ and\ \citenamefont
  {Nagai}}]{murakawa2009b}%
  \BibitemOpen
  \bibfield  {author} {\bibinfo {author} {\bibfnamefont {S.}~\bibnamefont
  {Murakawa}}, \bibinfo {author} {\bibfnamefont {Y.}~\bibnamefont {Wada}},
  \bibinfo {author} {\bibfnamefont {Y.}~\bibnamefont {Tamura}}, \bibinfo
  {author} {\bibfnamefont {M.}~\bibnamefont {Saitoh}}, \bibinfo {author}
  {\bibfnamefont {Y.}~\bibnamefont {Aoki}}, \bibinfo {author} {\bibfnamefont
  {R.}~\bibnamefont {Nomura}}, \bibinfo {author} {\bibfnamefont
  {Y.}~\bibnamefont {Okuda}}, \bibinfo {author} {\bibfnamefont
  {Y.}~\bibnamefont {Nagato}}, \bibinfo {author} {\bibfnamefont
  {M.}~\bibnamefont {Yamamoto}}, \bibinfo {author} {\bibfnamefont
  {S.}~\bibnamefont {Higashitani}}, \ and\ \bibinfo {author} {\bibfnamefont
  {K.}~\bibnamefont {Nagai}},\ }\href {\doibase 10.1088/1742-6596/150/3/032070}
  {\bibfield  {journal} {\bibinfo  {journal} {J. Phys.: Conf. Ser.}\ }\textbf
  {\bibinfo {volume} {150}},\ \bibinfo {pages} {032070} (\bibinfo {year}
  {2009}{\natexlab{b}})}\BibitemShut {NoStop}%
\bibitem [{\citenamefont {Murakawa}\ \emph {et~al.}(2011)\citenamefont
  {Murakawa}, \citenamefont {Wada}, \citenamefont {Tamura}, \citenamefont
  {Wasai}, \citenamefont {Saitoh}, \citenamefont {Aoki}, \citenamefont
  {Nomura}, \citenamefont {Okuda}, \citenamefont {Nagato}, \citenamefont
  {Yamamoto}, \citenamefont {Higashitani},\ and\ \citenamefont
  {Nagai}}]{murakawa2011}%
  \BibitemOpen
  \bibfield  {author} {\bibinfo {author} {\bibfnamefont {S.}~\bibnamefont
  {Murakawa}}, \bibinfo {author} {\bibfnamefont {Y.}~\bibnamefont {Wada}},
  \bibinfo {author} {\bibfnamefont {Y.}~\bibnamefont {Tamura}}, \bibinfo
  {author} {\bibfnamefont {M.}~\bibnamefont {Wasai}}, \bibinfo {author}
  {\bibfnamefont {M.}~\bibnamefont {Saitoh}}, \bibinfo {author} {\bibfnamefont
  {Y.}~\bibnamefont {Aoki}}, \bibinfo {author} {\bibfnamefont {R.}~\bibnamefont
  {Nomura}}, \bibinfo {author} {\bibfnamefont {Y.}~\bibnamefont {Okuda}},
  \bibinfo {author} {\bibfnamefont {Y.}~\bibnamefont {Nagato}}, \bibinfo
  {author} {\bibfnamefont {M.}~\bibnamefont {Yamamoto}}, \bibinfo {author}
  {\bibfnamefont {S.}~\bibnamefont {Higashitani}}, \ and\ \bibinfo {author}
  {\bibfnamefont {K.}~\bibnamefont {Nagai}},\ }\href {\doibase
  10.1143/JPSJ.80.013602} {\bibfield  {journal} {\bibinfo  {journal} {J. Phys.
  Soc. Jpn.}\ }\textbf {\bibinfo {volume} {80}},\ \bibinfo {pages} {013602}
  (\bibinfo {year} {2011})}\BibitemShut {NoStop}%
\bibitem [{\citenamefont {Choi}\ \emph {et~al.}(2006)\citenamefont {Choi},
  \citenamefont {Davis}, \citenamefont {Pollanen},\ and\ \citenamefont
  {Halperin}}]{choi2006}%
  \BibitemOpen
  \bibfield  {author} {\bibinfo {author} {\bibfnamefont {H.}~\bibnamefont
  {Choi}}, \bibinfo {author} {\bibfnamefont {J.~P.}\ \bibnamefont {Davis}},
  \bibinfo {author} {\bibfnamefont {J.}~\bibnamefont {Pollanen}}, \ and\
  \bibinfo {author} {\bibfnamefont {W.~P.}\ \bibnamefont {Halperin}},\ }\href
  {\doibase 10.1103/PhysRevLett.96.125301} {\bibfield  {journal} {\bibinfo
  {journal} {Phys. Rev. Lett.}\ }\textbf {\bibinfo {volume} {96}},\ \bibinfo
  {pages} {125301} (\bibinfo {year} {2006})}\BibitemShut {NoStop}%
\bibitem [{\citenamefont {Davis}\ \emph {et~al.}(2008)\citenamefont {Davis},
  \citenamefont {Pollanen}, \citenamefont {Choi}, \citenamefont {Sauls},
  \citenamefont {Halperin},\ and\ \citenamefont {Vorontsov}}]{davis2008}%
  \BibitemOpen
  \bibfield  {author} {\bibinfo {author} {\bibfnamefont {J.~P.}\ \bibnamefont
  {Davis}}, \bibinfo {author} {\bibfnamefont {J.}~\bibnamefont {Pollanen}},
  \bibinfo {author} {\bibfnamefont {H.}~\bibnamefont {Choi}}, \bibinfo {author}
  {\bibfnamefont {J.~A.}\ \bibnamefont {Sauls}}, \bibinfo {author}
  {\bibfnamefont {W.~P.}\ \bibnamefont {Halperin}}, \ and\ \bibinfo {author}
  {\bibfnamefont {A.~B.}\ \bibnamefont {Vorontsov}},\ }\href {\doibase
  10.1103/PhysRevLett.101.085301} {\bibfield  {journal} {\bibinfo  {journal}
  {Phys. Rev. Lett.}\ }\textbf {\bibinfo {volume} {101}},\ \bibinfo {pages}
  {085301} (\bibinfo {year} {2008})}\BibitemShut {NoStop}%
\bibitem [{\citenamefont {Mizushima}\ \emph {et~al.}(2012)\citenamefont
  {Mizushima}, \citenamefont {Sato},\ and\ \citenamefont
  {Machida}}]{mizushima2012}%
  \BibitemOpen
  \bibfield  {author} {\bibinfo {author} {\bibfnamefont {T.}~\bibnamefont
  {Mizushima}}, \bibinfo {author} {\bibfnamefont {M.}~\bibnamefont {Sato}}, \
  and\ \bibinfo {author} {\bibfnamefont {K.}~\bibnamefont {Machida}},\ }\href
  {\doibase 10.1103/PhysRevLett.109.165301} {\bibfield  {journal} {\bibinfo
  {journal} {Phys. Rev. Lett.}\ }\textbf {\bibinfo {volume} {109}},\ \bibinfo
  {pages} {165301} (\bibinfo {year} {2012})}\BibitemShut {NoStop}%
\bibitem [{\citenamefont {Ikegami}\ \emph {et~al.}(2013)\citenamefont
  {Ikegami}, \citenamefont {Tsutsumi},\ and\ \citenamefont
  {Kono}}]{ikegami2013}%
  \BibitemOpen
  \bibfield  {author} {\bibinfo {author} {\bibfnamefont {H.}~\bibnamefont
  {Ikegami}}, \bibinfo {author} {\bibfnamefont {Y.}~\bibnamefont {Tsutsumi}}, \
  and\ \bibinfo {author} {\bibfnamefont {K.}~\bibnamefont {Kono}},\ }\href
  {\doibase 10.1126/science.1236509} {\bibfield  {journal} {\bibinfo  {journal}
  {Science}\ }\textbf {\bibinfo {volume} {341}},\ \bibinfo {pages} {59}
  (\bibinfo {year} {2013})}\BibitemShut {NoStop}%
\end{thebibliography}
\end{document}